\documentclass[secnumarabic, footinbib,tightenlines,nobibnotes, aps, prb, twocolumn]{revtex4-1}
\usepackage{amsmath,amssymb}
\usepackage{graphicx}
\usepackage{dcolumn}
\usepackage{bm}
\usepackage{braket}
\usepackage{soul}
\usepackage{subcaption}
\usepackage{verbatim}
\usepackage{bbold}
\usepackage{color}
\usepackage{xcolor}
\usepackage{relsize}
\usepackage{amsthm}
\usepackage[T1]{fontenc}
\usepackage[colorlinks=true ,urlcolor=blue,urlbordercolor={0 1 1}]{hyperref}

\begin{document}

\title{Nearly Flat Chern Bands in Moir\'e Superlattices}
\author{
Ya-Hui Zhang}
\author{
Dan Mao
}
\author{Yuan Cao}
\author{ Pablo Jarillo-Herrero}
\author{
T. Senthil
}
\email{
senthil@mit.edu
}
\affiliation{Department of Physics, Massachusetts Institute of Technology, Cambridge, MA, USA
}

\date{\today}

\begin{abstract}
Topology and electron interactions are two  central themes in modern condensed matter physics. Here we  propose graphene based systems where both the band topology and interaction effects can be simply controlled with electric fields.   We study a number of systems of twisted double layers with small twist angle where a moir\'e super-lattice is formed. Each layer is chosen to be either  AB stacked bilayer graphene (BG), ABC stacked trilayer graphene (TG), or hexagonal boron nitride (h-BN).  In these systems  a vertical applied electric field enables control of the bandwidth, and interestingly also the  Chern number.    We find that the Chern numbers of the bands associated with each of the two microscopic valleys can be $\pm 0,\pm 1,\pm 2, \pm 3$ depending on the specific system and vertical electrical field.  We show  that these graphene moir\'e super-lattices are promising platforms to realize a number of fascinating many-body phenomena, including (Fractional) Quantum Anomalous Hall Effects.   We also discuss conceptual similarities and implications for modeling twisted bilayer graphene systems. 
\end{abstract}

\pacs{Valid PACS appear here}
\maketitle

\section{Introduction}
Van der Waals heterostructures have recently emerged as a highly tunable platform to study correlated electron physics. A striking recent example is the observation of correlated  insulating and superconducting states in twisted bilayer graphene encapsulated by hexagonal Boron Nitride (h-BN) near `magic' twist angles\cite{cao2018correlated,cao2018unconventional}.  The twisting produces a long period moir\'e pattern, and reconstructs the electronic band structure. Near the magic twist angle the  bands active near the Fermi energy become very  narrow\cite{bistritzer2011moire}. The effects of Coulomb interaction become important, and interesting correlation effects are seen.  Correlated insulating states have also been reported in trilayer graphene on  h-BN\cite{chen2018gate}. In other experiments magnetic field  induced fractional Chern insulators have been observed in moir\'e super-lattices of bilayer graphene\cite{spanton2018observation}. 
Further, an interesting recent  paper proposes to realize the triangular lattice Hubbard model in twisted heterostructures of transition metal dichalcogenides\cite{wu2018hubbard}. 

Here we describe graphene-based moir\'e super-lattice systems to realize model systems with two nearly flat bands which each carry equal and opposite Chern number $C$,   the most basic   topological index of a band\cite{thouless1982quantized}.   Previously  Ref. \onlinecite{song2015topological} demonstrated that when monolayer graphene is stacked on top of h-BN to form a moir\'e super-lattice, isolated $\pm C = 1 $ Chern bands  may potentially occur.   Other proposals for bands with  Chern number $\pm C$   in various graphene systems can be found  in Refs. \onlinecite{tong2017topological,san2014electronic,song2015topological,wolf2018substrate}. The width of these bands is large compared to the expected Coulomb energy and hence correlation effects may be expected to be weak.  Here we show that modifications of the proposal of Ref. \onlinecite{song2015topological} naturally enable engineering $\pm C \geq 1$ Chern bands which, moreover, are nearly flat. We dub such bands $\pm C $ Chern bands.  We find systems with various values of $C = 0, 1,2, 3$.    The near flatness  implies that correlation effects from Coulomb interaction are potentially  important.  Remarkably, $C$ can be controlled simply by applying a  transverse electric field.  Such an electric field also allows control of the bandwidth of the nearly flat band as shown experimentally\cite{chen2018gate}.  This allows electric control  of correlation effects, the band topology, and the band filling  and provides a possibly unique opportunity to study  many fascinating  phenomena.   Systems with flat Chern bands have also been studied theoretically for many years\cite{sheng2011fractional,neupert2011fractional,wang2011fractional,tang2011high,regnault2011fractional,sun2011nearly,bergholtz2013topological,neupert2015fractional,parameswaran2013fractional} as a potential realization of novel interacting topological phases of matter such as fractional Chern insulators.  Fractional Chern insulators   have been realized  in strong magnetic field experiments  with Harper-Hofstadter bands\cite{spanton2018observation} of bilayer graphene. However interacting topological phases  at zero magnetic field remain elusive in experiment.

As examples, we describe some natural and simple phases at  integer filings $\nu_T$ (which  is defined to be the number of electrons/moir\'e unit cell including spin and valley degrees of freedom) where,  we expect that the spin/valley degree of freedom will spontaneously uniformly polarize (`` spin/valley ferromagnet") so as to fully fill the band. For instance at $\nu_T = 1$ there is a spin and valley polarized insulating state which will show an electrical quantum Hall effect with $\sigma_{xy} = \frac{Ce^2}{h}$. This is an Integer Quantum Anomalous Hall (QAH) insulator.  QAH effects have been reported in magnetically doped three dimensional topological insulator(TI) systems\cite{chang2013experimental,checkelsky2014trajectory,kou2014scale,kandala2015giant}. However, QAH effect in a true two dimensional system with a Chern band has  never been achieved before.

At non-integer, but rational fillings $\nu_T$, a number of topological ordered states  - both Abelian and non-Abelian - seem possible. We  discuss some of these - specifically Fractional Quantum Anomalous Hall (FQAH) states - as well with a focus on states that have been previously established in numerical calculations on correlated flat Chern bands. Our  proposed realization of nearly flat $\pm C$ Chern bands with tunable interactions calls for a detailed exploration of the possible many body phases. We will mostly leave this for future work but remark that correlated $\pm C$ Chern bands may also be an excellent platform for the appearance of fractional topological insulators. 

For TG/h-BN correlated insulating states have already been reported in Ref.~\onlinecite{chen2018gate}. A model for one sign of the electric field when $C = 0$ was described in Ref.~\onlinecite{po2018origin}.  For the other sign of the electric field we get $\pm C = 3$ Chern bands, and we propose possible explanations of the observed insulating states in this side.

 Though the details are different, there is nevertheless a conceptual similarity between the Moire systems with $\pm C$ Chern bands discussed in this paper and the twisted bilayer graphene system. In particular, due to the non-zero Chern number of valley filtered bands, it is obvious that it is not possible to write down maximally localized Wanier functions while preserving the valley $U(1)$ symmetry as a local ``on-site'' symmetry.. Exactly the same feature appears - albeit in a more subtle form -  in the theoretical analysis of Ref.~\onlinecite{po2018origin} of the twisted bilayer graphene system. Thus both the systems discussed in this paper and the twisted bilayer graphene present an experimental context for a novel  theoretical problem where strong correlations combine with topological aspects of band structure. This makes their modeling  different from more conventional correlated solids.

\section{Models for Moir\'e Mini Band}
 At low energies the electronic physics of graphene is dominated by states near each of two valleys which we denote $\pm$. We start from a continuum model  of these low energy states:
\begin{equation}
  H_0=\sum_a\sum_{\mathbf k}\psi^\dagger_{a}(\mathbf k)h_{a}(\mathbf k)\psi_{a}(\mathbf k)
\end{equation}
where  $a=+,-$ is the valley index. $\psi_a(\mathbf k)$ is a multi-component vector with sublattice and layer index.

Time reversal transforms one valley to the other one, and requires that $h_+(\mathbf k)=h^{*}_{-}(-\mathbf k)$. In the following we focus on valley $+$. The model for valley $-$  can be easily generated by time reversal transformation.

 We  restrict attention to graphene with chiral stacking patterns, such as AB stacked bilayer graphene and ABC stacked trilayer graphene. The effective Hamiltonian for one valley $+$  of  such an $n$-layer graphene is
\begin{equation}
	h_{+}(\mathbf k)=\left(
	\begin{array}{cc}
	\frac{U}{2} & A(k_x-ik_y)^n \\
	A(k_x+ik_y)^n & -\frac{U}{2}
	\end{array}
	\right)
	\label{eq:h0}
\end{equation}
where $U$ is a mass term. $A=\frac{\upsilon^{n}}{\gamma_1^{n-1}}$, where $\upsilon \approx 10^6 m/s$ is the velocity of Dirac cone of each graphene layer and $\gamma_1\approx 380$ meV is the inter-layer coupling parameter. For monolayer graphene with $n=1$, $U$ is the energy difference between two sublattices. For AB stacked bilayer graphene (BG) with $n=2$ and ABC stacked trilayer graphene (TG) with $n=3$, $U$ is the energy difference between top layer and bottom layer. Therefore $U$ can be easily controlled by perpendicular electric field in BG and TG, which are the main focus in this paper.

 When two lattices with lattice constant $a_1$ and $a_2$ are stacked with a small twist angle $\theta$ or when they have slightly different lattice constants with $\xi=\frac{a_1-a_2}{a_2}$, there is a large moir\'e super-lattice with lattice constant $a_M=\frac{a}{\sqrt{\xi^2+\theta^2}}$ at small twist angle $\theta$, where $a=\frac{a_1+a_2}{2}$.  The moir\'e lattice reconstructs the original band into  a small Moir\'e Brillouin Zone(MBZ) which is a hexagon with size $|K|=\frac{4\pi}{3 a_M}$. The calculation of band structure\cite{bistritzer2011moire} is very similar to the classic example of  free electron approximation in elementary solid state textbooks\cite{Ashcroft}. Details can be found in the Appendix.~\ref{appendix:methods}.   When $a_M \gg a$, valley mixing terms can be ignored, and two valleys can be treated separately.

\onecolumngrid

\begin{figure}[ht]
\centering
  \includegraphics[width=\textwidth]{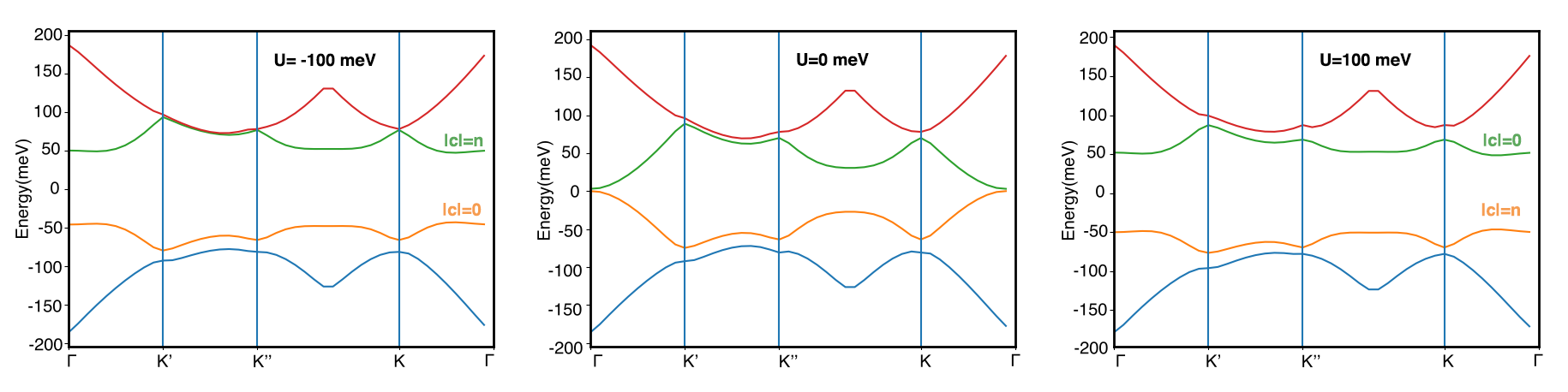}
  \caption{Possibility of gate driven nearly flat  Chern band for twisted graphene/h-BN systems. $U$ is the potential difference between the top and bottom layers, controlled by applied vertical electric field. Figures are generated for BG/h-BN. TG/h-BN shows quite similar behavior. $n=2$ for BG/h-BN and $n=3$ for TG/h-BN. Two sides with opposite $U$ have similar band dispersions with Chern number $|C|=0$ and $|C|=n$ respectively, which enables  to switch between Hubbard model physics and quantum Hall physics within one sample.  Twist angle $\theta=0$ and $a_M\approx 15$ nm. $K'=(0,\frac{4\pi}{3a_M})$. $K''=(\frac{2\pi}{\sqrt{3}a_M},\frac{2\pi}{3a_M})$ and $K=(0,-\frac{4\pi}{3a_M})$ are equivalent in the MBZ.}
  \label{fig:main_text_gate_driven_flat_band}
\end{figure}

\twocolumngrid

We discuss two different categories of moir\'e systems in this article. (1) $n$-layer graphene on top of h-BN with small twist angle.  Each valley maps to the $\Gamma$ point of the MBZ. (2) $n_1$-layer graphene on top of  $n_2$-layer graphene  with a relative twist angle $\theta$ close to a `magic' value $\approx 1^\circ$. In the MBZ, the valley $+$ of top and bottom graphene layers map to the $K$  and  $K'$ points {}respectively. Valley $-$ is related by time reversal transformation.  We assume that the two graphenes have the same chirality of stacking pattern. For example, we assume both BG are AB stacked in the BG/BG system.  Systems with opposite chirality are discussed in the Appendix.~\ref{appendix:graphene_graphene}.

In these systems, generically only $C_3$ rotation symmetry is preserved while  the inversion symmetry or equivalently $C_6$ rotation symmetry is broken. Time reversal symmetry flips the valley index and  only requires that the Chern numbers for the bands of two valleys are opposite. 

In the next two sections we  demonstrate that these systems show the following two interesting features : (1) Nearly flat bands; (2)Non-zero Chern number of the narrow bands for each valley.  The small bandwidth    implies that strong correlation effects may be present and play an important role in determining the physics.  The non-zero  $\pm$ Chern number further  renders the physics different from traditional Hubbard-like lattice models  which are only appropriate for  topologically trivial bands.

\section{Nearly Flat Bands}
\subsection{Twisted Graphene/h-BN Systems}

\begin{figure}[ht]
  \centering
    \includegraphics[width=0.45\textwidth]{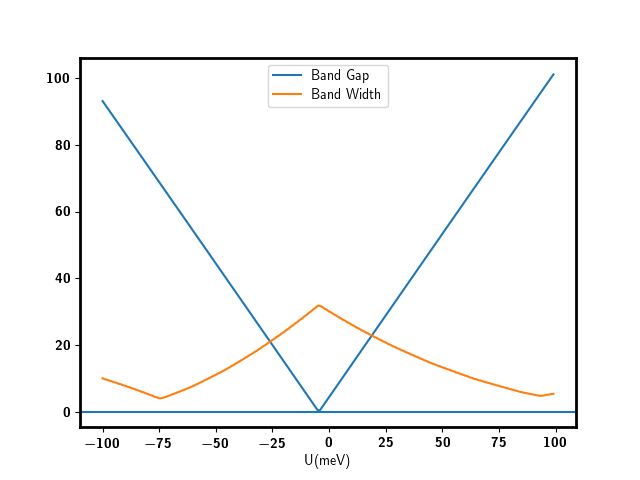}
    \caption{Bandwidth of valence band with the applied voltage $U$ for the TG/h-BN system.}
  \label{fig:main_text_tg_bn_band_width}
  \end{figure}

As shown experimentally for the TG/h-BN system in Ref.~\onlinecite{chen2018gate}, the width of  the valence and  conduction bands can be tuned by applying perpendicular  electric field $\propto U$ (please see  Fig.~\ref{fig:main_text_tg_bn_band_width}). We  further demonstrate this theoretically for the BG/h-BN system  in Fig.~\ref{fig:main_text_gate_driven_flat_band}.

\subsection{Twisted Graphene/Graphene Systems}
For twisted graphene/graphene systems, there exist  ``magic angles" at which the  bandwidth is reduced to almost zero. This  demonstrates  that the existence of  magic angles  is a  common feature for graphene/graphene systems and  is not special to the twisted bilayer graphene system studied  in Ref.~\onlinecite{bistritzer2011moire}. Given the discovery of  correlated insulating and superconducting states in twisted bilayer graphene, we expect these  other systems will also show interesting  correlation driven physics.

For BG/BG and TG/BG system, the magic  angle is the same as the twisted bilayer graphene: $\theta_M=1.08^\circ$. For TG/TG, we have $\theta_M=1.24^\circ$. A plot of bandwidth vs  twist  angle is shown in Fig.~\ref{fig:main_text_magic_BG_BG} for the BG/BG system. Details for the other two systems can be found in  Appendix.~\ref{appendix:graphene_graphene}).

\begin{figure}
	\centering
    \includegraphics[width=0.45\textwidth]{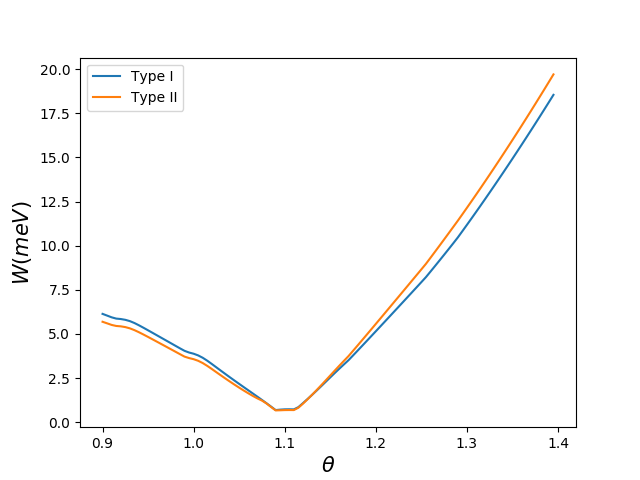}
    \caption{Bandwidth of conduction band with twist angle for the BG/BG system.}
	\label{fig:main_text_magic_BG_BG}
  \end{figure}

\section{Analysis of Chern number for Moir\'e mini band}

Our results for Chern numbers at small $|U|$ are summarized in Table.~\ref{table:chern_number}.  We only show  the absolute value $|C|$ of the Chern number because the two valleys always have opposite Chern numbers. This table is  one of the key results of this article.  Below we explain these results.

 For $n$-layer graphene/h-BN systems,  the  ``mass" term $U$ in Eq.~\ref{eq:h0} gaps out  the band crossing at the $\Gamma$ point, as shown in Fig.~\ref{fig:main_text_gate_driven_flat_band}. Typically different directions of perpendicular electric field give very similar band structure.  When  the sign of $U$  changes, there is a band inversion which changes the  contribution to the Berry curvature from states near the band touching points. In ordinary $n$-layer graphene the two valleys are connected in momentum space and their opposite Berry curvatures implies that the  net  Chern number is zero. The moir\'e potential however detaches the minibands in each valley from the rest of the spectrum, and their individual Chern numbers $\pm C$ become well defined.  The exact value of $C$ of each miniband depends on contributions away from the band touching points, which do not change with sign of $U$.  For $n$-layer graphene,  Eq.~\ref{eq:h0} readily gives  $\Delta C = C(U) - C(-U) = n$. Detailed calculations show that $C(U<0)=0$ and $C(U>0)=n$ for the valence band.  We have verified that the Chern number is quite stable to different parameters used for the moir\'e super-lattice potential.

\begin{table}
\begin{center}
\begin{tabular}{ c|c|c } 
 \hline
 Systems & $U<-\Delta_0$ & $U>\Delta_0$ \\ 
 \hline
 BG/h-BN,valence & $0$ & $2$ \\ 
 \hline
 TG/h-BN,valence & $0$ & $3$ \\ 
 \hline
 BG/h-BN,conduction & $2$ & $0$ \\ 
 \hline
 TG/h-BN,conduction & $3$ & $0$ \\ 
 \hline
 \hline
 BG/BG &  $2$ & $2$  \\ 
 \hline
 TG/TG &  $3$ & $3 $  \\ 
 \hline
 TG/BG, conduction &  $2$  & $3$  \\ 
 \hline
 TG/BG, valence &  $3$  & $2$  \\ 
 \hline
\end{tabular}
\end{center}
\caption{Chern number $|C|$ of conduction/valence bands for twisting graphene/h-BN ($\Delta_0=0$)  and graphene/graphene systems ($\Delta_0\lessapprox2$ meV).}
\label{table:chern_number}
\end{table}

For the graphene/graphene system with $n_1$ and $n_2$ layers, we found that there is already a small gap $\Delta_0\sim 2$ meV for the band crossing point at $K$ and $K'$ point. We only discuss the case $|U|>\Delta_0$. With the same argument as in the graphene/h-BN system, we get $\Delta C=n_1+n_2$ for the valence band when  $U$ goes from negative to positive. For the BG/BG and TG/TG systems with $n_1=n_2=n$, there is a mirror reflection symmetry which requires $C(U)=-C(-U)$. Then we can show analytically that $C=\pm n$. For the TG/BG system, we rely on numerical calculations to get the results in Table.~\ref{table:chern_number}.

\begin{figure}[ht]
\centering
 \includegraphics[width=0.45\textwidth]{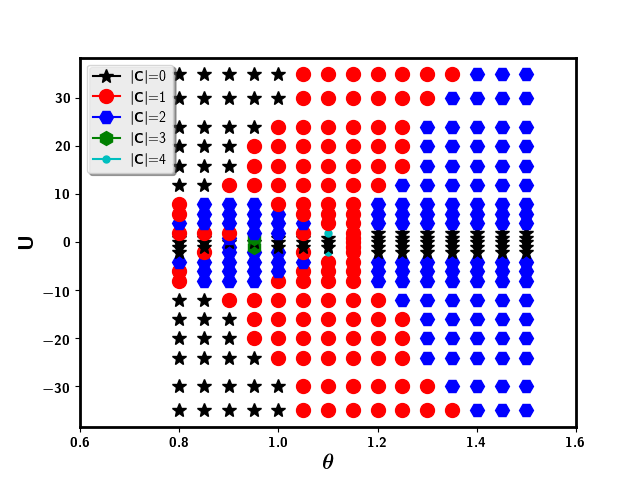}
   
  \caption{Chern number $|C|$ of BG/BG for conduction band. $U$ is in the unit of meV.  There is a clear phase boundary $U_c(\theta)$, at which Chern number jumps from $|C|=2$ to $|C|=1$ because the hybridization gap between conduction band and the band above closes at $\Gamma$ point.     $U_c(\theta_M)=0$ at magic angle $\theta_M$ in our simple model. As explained in the main text, our method may underestimate $U_c(\theta)$ and $U_c(\theta)$ may remain large even at magic angle in realistic systems. }
  \label{fig:abs_C_Plot_BG_BG}
\end{figure}

The results above are based on the assumption that $|U|$ is small. For graphene/graphene systems, there is a phase boundary $U_c(\theta)$ across which the Chern number $|C|$ drops by $1$.  The phase diagram for the  BG/BG system is shown in Fig.~\ref{fig:abs_C_Plot_BG_BG}. $U_c(\theta)$ goes to almost zero at the  magic  angle $\theta_M$ in our simple continuum model. At $U_c$, the gap between the valence band and the band below  closes.  However, it is likely that lattice relaxation and interaction effects  will  enhance this hybridization gap and as a consequence $U_c(\theta_M)$ may remain large even at the magic angle in realistic experimental systems. We leave it to experiment to measure the true hybridization gap\footnote{For a clean sample, the valley contrasting Chern number may also be extracted from the conductance of the helical edge modes in the fully filled Quantum Valley Hall insulator at $\nu_T=4$.}.

\section{Hamiltonian of Chern Bands \label{section:hamiltonian}}

 Having shown that nearly flat Chern bands are possible in graphene based moir\'e systems we now discuss the Hamiltonian to model these $\pm$ Chern bands. The model of course needs to have a global $U_c(1)$ charge conservation symmetry, and spin $SU(2)$ rotation symmetry. In addition the valley   $U_v(1)$ symmetry is an excellent approximate symmetry and we will treat it as exact.

 The {\em microscopic} Hamiltonian is:
\begin{equation}
  H=H_{band}+H_{int}
\end{equation}
 where $H_{band} $ is the band dispersion and $H_{int}$ is the electron-electron interaction:

 \begin{equation}
  H_{int} =\sum_{a_1,a_2;\sigma_1,\sigma_2}\sum_{\mathbf q} \rho_{a_1,\sigma_1}(\mathbf q) V(\mathbf q)\rho_{a_2,\sigma_2}(-\mathbf{q})
  \label{eq:chern_interactionbare}
\end{equation}
where summation of $\mathbf{q}$ is in the whole $R^2$ space instead of the Mini Brillouin Zone.   We use the Coulomb potential $V(\mathbf q)\sim  \frac{1}{|\mathbf q|}$.

A  low energy  Hamiltonian is obtained by projecting this microscopic Hamiltonian to the active bands near the Fermi energy. The kinetic term is then restricted to 
\begin{equation}
  H_0=\sum_{a\sigma}\sum_k c^\dagger_{a\sigma}(k)\xi_a(k) c_{a\sigma}(k) 
  \label{hkin}
\end{equation}
where $a=+,-$ is valley index and $\sigma=\uparrow,\downarrow$ is spin index. $\xi_+(k)=\xi_-(-k)$ is guaranteed by time reversal symmetry.

 The {\em leading term} in the interaction part of the effective Hamiltonian will again be the density-density repulsion but we must use density operators projected to the active bands.  For concreteness we will describe this below for the valence band. 
Valence band annihilation operator $c(\mathbf k)$ can be expressed in terms of the microscopic electron operator $f_{\alpha;a,\sigma}(\mathbf k)$:
\begin{equation}
  c_{a,\sigma}(\mathbf k)=\sum_{m,n} \mu_{mn,\alpha;a}(\mathbf k)f_{\alpha;a,\sigma}(\mathbf k+m\mathbf{G_1}+n \mathbf{G_2})
  \label{eq:bloch}
\end{equation}
where $\mu_{mn,\alpha;a}(\mathbf k)$ is the Bloch wave function of valence band. $m,n$ is the label of the momentum points of $M\times M$ grid in free electron approximation. $\alpha$ is label of orbital for the microscopic electron $f_{\alpha;a,\sigma}(k)$, for example $\alpha=t,b$ labels top layer and bottom layer of BG.  
In terms of $f_{\alpha;a,\sigma}(\mathbf k)$, we have $\rho_a(\mathbf q)=\sum_{\sigma} f^\dagger_{\alpha;a,\sigma}(\mathbf{k+q})f_{\alpha;a,\sigma}(\mathbf k)$. We  emphasize that here we distinguish $\mathbf q$ and $\mathbf q+m \mathbf{G_1}+ n \mathbf{G_2}$ as different momenta.

From Eq.~\ref{eq:bloch} we  get:
\begin{equation}
  f_{a;\alpha}(\mathbf k+m\mathbf{G_1}+n \mathbf{G_2})=\mu^*_{mn,\alpha;a}(\mathbf k) c_a(\mathbf k)+...
\end{equation}
where $...$ denotes other bands and we have suppressed spin index.

 We can now represent the projected density operator $\tilde{\rho}_a(\mathbf q)$ in terms of $c_a(\mathbf k)$:
\begin{equation}
 \tilde{\rho}_{a,\sigma}(\mathbf{q})=\sum_{\mathbf k} \lambda_a(\mathbf k, \mathbf{k+q}) c^\dagger_{a,\sigma}(P(\mathbf k + \mathbf q))c_{a,\sigma}(\mathbf k)
  \label{eq:form_factor}
\end{equation}
where $P(\mathbf k )=\mathbf k_0$  projects $\mathbf k$ to $\mathbf k_0$ in the Moir\'e Brillouin Zone(MBZ) if $\mathbf k=\mathbf k_0+m \mathbf{G_1}+n\mathbf{G_2}$.

 The form factor $\lambda_a$ is defined through the  Bloch wave function $\mu_a(\mathbf k)$:
\begin{equation}
  \lambda_a(\mathbf k,\mathbf{k+q})=\braket{\mu_a(\mathbf k)|\mu_a(\mathbf{k+q})}
\end{equation}

If $\mathbf k=\mathbf k_0+m_0 \mathbf{G_1}+n_0\mathbf{G_2}$ is not in the MBZ, $\mu_{mn,\alpha;a}(\mathbf k)$ is defined as:
\begin{equation}
  \mu_{m,n,\alpha;a}(\mathbf k)=\mu_{m+m_0,n+n_0,\alpha;a}(\mathbf{k_0})
\end{equation}

We emphasize again that $V_{a_1,a_2}(\mathbf q+ m \mathbf{G_1} +n \mathbf {G_2})\neq V_{a_1,a_2}(\mathbf q)$ unless $m,n=0$. Correspondingly $\lambda_a(\mathbf k,\mathbf{k+q})\neq \lambda_a(\mathbf k,\mathbf{k+q+m G_1+n G_2})$ unless $m,n=0$. This is the reason  we need to sum over $\mathbf q \in R^2$ in Eq.~\ref{eq:chern_interaction}. In practice one can always truncate $\mathbf {q}=P(\mathbf{q})+m \mathbf G_1+n \mathbf {G_2}$ with $|m|,|n|<M$. $M=1$  corresponds to summation over only MBZ. 

$c_a(k)$, $\mu_a(\mathbf k)$ and $\lambda_a(\mathbf{k},\mathbf{k+q})$ are all not gauge invariant. There is a gauge degree of freedom:
\begin{align}
c_a(k)&\rightarrow c_a(k)e^{i\theta_a(k)}\notag\\
\mu_a(\mathbf k)&\rightarrow \mu_a(\mathbf k)e^{i\theta_a(k)}\notag\\
\lambda_a(\mathbf k, \mathbf{k+q})&\rightarrow \lambda_a(\mathbf k, \mathbf{k+q})e^{i\left(\theta_a(\mathbf{k+q})-\theta_a(\mathbf k)\right)}
\end{align}

One can easily check that $\tilde{\rho}_a(\mathbf q)$ is gauge invariant. For each valley $a$, if the band is  topologically trivial, we can fix the gauge $\theta_a(\mathbf k)$ globally. However, if the band for this valley has non-zero Chern number, there is no way to fix a global gauge in the whole Brillouin zone.  For this situation,  the density operator $\tilde{\rho}_a(\mathbf q)$ is shown to satisfy Girvin-MacDonald-Platzman algebra\cite{girvin1986magneto} (also called $W_\infty$ algebra) in the $q\rightarrow 0$ limit\cite{GMP-Sondhi}, which is familiar from the physics of a single  Landau level for quantum hall systems  in a high magnetic field. 

 The effective density-density interaction takes the form 
\begin{equation}
  H_V  =\sum_{a_1,a_2;\sigma_1,\sigma_2}\sum_{\mathbf q} \tilde{\rho}_{a_1,\sigma_1}(\mathbf q) V(\mathbf q)\tilde{\rho}_{a_2,\sigma_2}(-\mathbf{q})
  \label{eq:chern_interaction}
\end{equation}
where summation of $\mathbf{q}$ is in the whole $R^2$ space instead of the Moir\'e Brillouin Zone.   We use the Coulomb potential $V(\mathbf q)\sim  \frac{1}{|\mathbf q|}$.  

Due to the non-trivial algebra satisfied by the $\tilde{\rho}$ operators, interaction effects are qualitatively different from the case with zero Chern number.  Actually, it has been shown that for a single Chern band, interactions may drive the system into exotic  topological ordered states similar to Fractional Quantum Hall states\cite{parameswaran2013fractional}. We will address this possibility in  subsequent sections.

The terms $H_0 + H_V$ (from Eqns. \ref{hkin} and \ref{eq:chern_interaction}) define an approximate effective Hamiltonian for the partially filed valence band.  At this level of approximation, the electron charge and the electron spin in each valley is separately conserved.   With just these terms the effective Hamiltonian thus has a $U(2) \times U(2)$ symmetry. This will be further broken down to the assumed physical symmetries of just charge $U_c(1)$, valley $U_v(1)$, and total spin $SU(2)$ by weaker terms. The pertinent such weaker interaction is a Hund's coupling:
\begin{equation}
H_J = - J_H  \int d^2 x \sum_{a_1,a_2} \mathbf{S}_{a_1}(\mathbf q) \cdot \mathbf{S}_{a_2}(-\mathbf q)
\label{eq:Hund}
\end{equation}
where $\mathbf{S}_a(q)$ is the spin operator associated with electrons in valley $a$ at a momentum $q$. Like the density operator itself $S_a(q)$ is also projected to the valence band and takes the form 
\begin{equation}
\mathbf{S}_a(q) = \frac{1}{2}\sum_{\mathbf k; \sigma_1\sigma_2} \lambda_a(\mathbf k, \mathbf{k+q}) c^\dagger_{a,\sigma_1}(P(\mathbf k + \mathbf q)\mathbf {\sigma_{\sigma_1\sigma_2}} c_{a,\sigma_2}(\mathbf k)
\end{equation}

We expect that $J_H$ is weaker than the Coulomb scale by roughly a factor $\frac{a}{a_M}$. We will therefore initially ignore it and will include its effects as needed later.

 For completeness we note that we can also define the valley density, or more generally, a $SU(2)_v$ ``valley pseudo-spin" operator projected to the valence band:
\begin{equation}
  I_{\mu}(\mathbf{q})=\sum_{ab;\mathbf k\sigma} \lambda_{ab}(\mathbf k, \mathbf{k+q}) c^\dagger_{a,\sigma}(P(\mathbf k + \mathbf q))\tau^{\mu}_{ab}c_{b,\sigma}(\mathbf k)
  \label{eq:valley_operator}
\end{equation}
 where,
 \begin{equation}
 	\lambda_{ab}(\mathbf k, \mathbf{k+q})=\braket{\mu_a(\mathbf k)|\mu_b(\mathbf{k+q})}
 \end{equation}
 Only the $I_z (q = 0) $ operator is conserved. 
 
 Our final approximate effective Hamiltonian for the partially filled valence band takes the form
 \begin{equation}
 H_{eff} = H_0 + H_V + H_J
 \end{equation}
with each of the three terms on the right given by Eqns. \ref{hkin}, \ref {eq:chern_interaction}, and \ref{eq:Hund} respectively.

\section{Quantum Anomalous Hall Effect and Quantum Valley Hall Effect}
We now discuss the possible insulating phases at integer filling $\nu_T$ of valence bands with non-zero Chern number. $\nu_T$ is defined as the total density for each moir\'e unit cell. $\nu_T=4$ corresponds to filling four bands completely.  Here we only focus on the case $|C|\neq 0$  and $\nu_T = 1, 2, 3$. For  $C=0$, there is  a localized Wannier orbital for each valley and the physics is governed by an extended Hubbard model. The $C=0$ side of the ABC trilayer graphene/h-BN system is discussed in details elsewhere\cite{zhang2018bridging}. In this paper we focus on the topologically non-trivial side.  For the $C\neq 0$ side, Wannier localization is not possible for each valley.
Due to the Wannier obstruction, a simple  Hubbard-like  lattice model  is impossible. Physically  we can not think of  correlated insulating phases at partial band filling  in the traditional ``charge frozen" picture of a  Hubbard model.  The systems discussed in the current paper thus pose the novel theoretical problem of correlation effects in a partially filled topological band.  A  general framework to deal with this kind of system is currently absent. If we initially ignore $H_J$ the model has two energy scales: $V$ for the Coulomb interaction and $W$ for the band width.  Here we restrict ourselves to the strong interaction limit  $\frac{V}{W}\rightarrow 0$.  Guided by previous studies on multi-component  quantum hall systems,  we propose that the insulating phases at $\nu_T = 1,2,3$  are ``ferromagnetic"  fully polarized in spin-valley space so as to spontaneously yield completely filled bands.   Hartree Fock calculation is enough to get the ground state. Below we will  explore this proposal within a simple Hartree-Fock calculation.

\subsection{Hartree Fock Calculation of Quantum Hall Ferromagnetism}
Similar to quantum Hall systems of graphene in a  high magnetic filed,  in the flat band limit, Coulomb interaction tends to drive the system to uniformly polarize spin or valley and fill band of the same species. Because of full $SU(2)$ spin rotation symmetry, different directions of spin polarization are degenerate.  The valley degree of freedom only has $U(1)$ rotation symmetry in the $z$ direction.  Then we need to decide whether $I_z$ or $I_x$ (or equivalently $I_y$ because of the rotation symmetry generated by $I_z$) is polarized. In the remaining part of this subsection we perform  a Hartree Fock  calculation  to shows that the system prefers to polarize the valley in the $z$ direction instead of in the $xy$ direction.

 We consider the general repulsive interaction in Eqn. \ref{ eq:chern_interaction}, and study the Hartree-Fock energy of fully polarized (in spin-valley space) states at integer $\nu_T$.  We assume that $V \gg W$ but do not simply take the exact flat band limit $W = 0$. 
 The  Hartree energy is always the same for the state with the same density. The only difference comes from the Fock exchange energy:
\begin{align}
  H_F&=-\sum_{a_1\sigma_1;a_2\sigma_2;\mathbf{k},\mathbf{q}}V(\mathbf{q})\lambda_{a_1}(k,q)\lambda_{a_2}(k+q,-q)\notag\\
  & \langle c^\dagger_{a_1,\sigma_1}(\mathbf{k+q})c_{a_2;\sigma_2}(\mathbf{k+q})\rangle \langle c^\dagger_{a_2,\sigma_2}(\mathbf k)c_{a_1\sigma_1}(\mathbf k)\rangle
\end{align}

First we consider a Hartree Fock state which does not mix different $a_1,\sigma_1$ and $a_2,\sigma_2$:
\begin{equation}
  \langle c^\dagger_{a_1\sigma_1}(k) c_{a_2 \sigma_2}(k) \rangle=\delta_{a_1 a_2}\delta_{\sigma_1 \sigma_2} F_{a_1 \sigma_1}(\mathbf k)
\end{equation}
where $F_{a\sigma}(\mathbf k)=\theta(\mu-\xi_a(\mathbf k))$ is Fermi-Dirac distribution at zero temperature limit. If the fermi energy contour is a circle, we can use $F_{a\sigma}(\mathbf k)=\theta(k_F-k)$. We will assume this in the following. For the case where the full band is filled satisfied for integer $\nu_T$, this assumption trivially holds and we can simply remove $F_{a\sigma}(\mathbf k)$ factor.  

For a state for which valley $a$ is polarized\footnote{ Such a state can be thought as the ground state of an $I_z$ polarized mean field Hamiltonian: $H_M=H_K-M I_z(\mathbf q=0)$ with $M\rightarrow \infty$)}, the Fock exchange energy is simplified to:
\begin{equation}
   E_F=-\sum_{\mathbf{k},\mathbf{q}}V(\mathbf q)|\lambda_{a}(\mathbf{k},\mathbf{q})|^2
\end{equation}
where we have use $\lambda(k,q)=\lambda(k+q,-q)^*$.

The kinetic energy is
\begin{equation}
  E_{K;0}= \sum_{\mathbf k} \xi_{a\sigma}(\mathbf k)
  \label{eq:kinetic_polarized}
\end{equation}

Next we also calculate the energy of an inter-valley-coherent (IVC) state with full valley polarization in the $I_x$ direction \footnote{ Such a state can be viewed as the ground state of the following  $I_x$ polarized mean field Hamiltonian $H_M=H_K-M I_x(\mathbf q=0)$ with $M\rightarrow + \infty$.}.

Using Eq.~\ref{eq:valley_operator}, the IVC state can be written:
\begin{equation}
  \ket{\Psi_{IVC};\sigma}=\prod_{|k|<k_F}\frac{1}{\sqrt{2}}(e^{i\theta_{+}(k)} c^\dagger_{+;\sigma}(k)+e^{i\theta_{-}(k)}c^\dagger_{-;\sigma}(k))\ket{0}
\end{equation}
where $\theta_{a}(\mathbf k)$ is the phase of the Bloch-wavefunction $\mu_{a}(\mathbf k)$. The above state is independent of the gauge choice for the Bloch wavefunction.

Then we calculate the expectation value of energy:
\begin{equation}
  E_{IVC}=\braket{\psi_{IVC}|H_K+H_V|\psi_{IVC}}
\end{equation}

For simplicity we drop out the spin index. The kinetic energy is straightforward:
\begin{equation}
  E_K=\sum_k \frac{1}{2}(\xi_{+}(\mathbf k)+\xi_{-}(\mathbf k))=\sum_{\mathbf k} \xi_{+}(\mathbf k)
\end{equation}
which is the same as the kinetic energy of an $I_z$ polarized state.

The Fock energy for the IVC state is:
\begin{align}
E_{F;IVC}=&-\frac{1}{4}\sum_{a_1,a_2;k,q}\big(V(\mathbf{q})
\lambda_{a_1}(\mathbf k, \mathbf q)\lambda_{a_2}(\mathbf{k+q},-\mathbf q) \notag\\
&e^{-i(\theta_{a_1}(\mathbf{k+q})-\theta_{a_1}(\mathbf{k}))}e^{i(\theta_{a_2}(\mathbf{k+q})-\theta_{a_2}(\mathbf k))}\notag\\
=&-\frac{1}{4}\sum_{a_1,a_2;k,q}\big(V(\mathbf{q})
|\lambda_{a_1}(\mathbf k,\mathbf q)||\lambda_{a_2}(\mathbf k,\mathbf q)|\notag\\
\end{align}
where the phase of $\lambda_{a}(\mathbf{k},\mathbf{q})$ is removed by the additional phase factor $e^{i\theta_a(\mathbf k)}$. As expected, the final expression does not depend on the phase of $\lambda_a(\mathbf k, \mathbf q)$, which is not gauge invariant. We have used the identity $\lambda_a(\mathbf{k+q},-\mathbf q)=\lambda_a^*(\mathbf k,\mathbf q)$.

The Fock energy from the $I_z$ polarized  state is
\begin{align}
E_{F;0}&=-\sum_{\mathbf k,q}V(\mathbf q) |\lambda_{+}(\mathbf k, \mathbf q)|^2\notag\\
&=-\frac{1}{4}\sum_{k,q} V(\mathbf q)\big(2|\tilde \lambda_{+}(\mathbf k,\mathbf q)|^2+2|\tilde \lambda_{-}(\mathbf k,\mathbf q)|^2\big)
\end{align}

Finally, we can get the energy difference between IVC state  and the $I_z$ polarized state:
\begin{align}
  \Delta E_{IVC}&=E_{F;IVC}-E_{F;0}+E_{K;IVC}-E_{K;0} \notag\\
  &=\frac{1}{4}\sum_{\mathbf k, \mathbf q}V(\mathbf q)\big(|\lambda_+(\mathbf{k},\mathbf{q})|-|\lambda_{-}(\mathbf{k},\mathbf{q})|\big)^2\notag\\
  >0
\end{align}

Time reversal requires $|\lambda_{+}(\mathbf{k},\mathbf{q})|=|\lambda_{-}(-\mathbf{k},-\mathbf{q})|$. However, generically $|\lambda_{+}(\mathbf{k},\mathbf{q})|$ is not equal to $|\lambda_{-}(\mathbf{k},\mathbf{q})|$ because inversion is broken for each valley.  Therefore $I_z$ polarization is always favored in the flat band limit for repulsive interaction $V(\mathbf q)>0$. This conclusion is independent of the interaction used (as long as it is repulsive) and the details of the band structures.  It is generically true for all of the spin-valley bands with opposite Chern numbers for the two valleys in the flat band limit.  Once the bandwidth is large, the IVC order ($I_x$ order) may indeed be favored compared to the $I_z$ polarization because of the kinetic terms, as discussed in Ref.~\onlinecite{po2018origin}. In this paper we focus on the limit that bandwidth is much smaller than the interaction. Then we will only discuss the possibility of the $I_z$ polarization.

\hfill \break

 Next we discuss the possible states for different integer fillings based on the favored $I_z$ polarization from the Hartree Fock calculation. The TG/h-BN system has already been shown to have correlated insulators in $\nu_T=1$ and $\nu_T=2$\cite{chen2018gate}. The following discussions can also apply to TG/h-BN system for one direction of the vertical displacement field. 

At $\nu_T=1$, we expect one band is filled completely. Let us assume that it is the band $(+,\uparrow)$.  For the next band,  density density interaction gives the same energy to spin polarized state and valley polarized state.  Including the Hund's coupling, we expect spin polarized state is selected. Therefore  we expect that the next band to be filled after $\nu_T=1$ is $(-,\uparrow)$.  The third and the fourth band to fill are either $(-,\downarrow)$, $(+,\downarrow)$ or $(+,\downarrow)$, $(-,\downarrow)$. These two choices do not make too much difference  for the properties of the insulator at filling $\nu_T=3$.  We therefore assume the four bands are filled in the following sequence: $(+,\uparrow)$, $(-,\uparrow)$, $(+,\downarrow)$, $(-,\downarrow)$. This sequence is only meaningful at integer $\nu_T$. For fractional filling of $\nu_T$, spin singlet or valley singlet states are possible, which depend on energetics not captured by the simple Hartree Fock calculation.

 Below we will discuss the property of insulators at these filling  $\nu_T=1,2,3$.

\subsection{$\nu_T=4$: Quantum Valley Hall Insulator with $\sigma_{xy}^\upsilon=4C \frac{e^2}{h}$}
We start with the trivial case  $\nu_T=4$ when we get a band insulator from fully filling all four  bands . It has quantized integer ``valley hall conductivity" $\sigma_{xy}^v=4 C \frac{e^2}{h}$. If $U_v(1)$ valley symmetry is preserved at the sample edge, there are $4C$ number of helical edge modes: $2C$  edge modes of one valley  moving in one direction and  $2C$ edge modes of the other valley moving in the opposite direction. Ideally, there should be a quantized conductance $G=4C \frac{e^2}{h}$ coming from these $4C$ edge channels. However, impurities at the edge may back-scatter the two different valley modes  and hence reduce the conductance. To reduce valley scattering, it may be easier to observe these edge modes at the  domain wall between AB (ABC) stacked and BA (CBA) stacked BG (TG) on h-BN systems. It has been shown experimentally that such a smooth domain wall can preserve the valley index\cite{ju2015topological}. For BG/h-BN and TG/h-BN system, at the domain wall (as opposed to the sample edge), we expect to observe $4C$ edge modes for one valley and $4C$ edge modes going in the opposite direction for the other valley.

\subsection{$\nu_T=2$: Spin polarized Quantum Valley Hall Insulator with $\sigma_{xy}^\upsilon=2C \frac{e^2}{h}$}

At $\nu_T=2$,  the spin-valley polarized state will have  two filled bands. We expect that the Hund coupling selects the spin polarized state. Then we have a quantum valley hall insulator with $\sigma_{xy}^\upsilon=2C \frac{e^2}{h}$. Properties are quite similar to $\nu_T=4$ IQVHE effect, except that the number of  helical edge modes is $C$ (one half of the $\nu_T=4$ case). Besides, as usual, there will be a skyrmion defect of the ferromagnetic order parameter that carries a valley charge $Q_\upsilon=2C$ but is electrically neutral.

\subsection{$\nu_T=1$ and $\nu_T=3$: Quantum Anomalous Hall Insulator with $\sigma_{xy}^c=C \frac{e^2}{h}$}
For  $\nu_T=1$ and $\nu_T=3$,   valley and spin are both polarized to form a filled band. Then there is an IQHE effect with Hall conductivity $\sigma_{xy}^\upsilon=C \frac{e^2}{h}$.  This is a Quantum Anomalous Hall insulator, and depending on the system, has Chern number $C > 1$ for which there is no previous experimental realization in a solid state system. At $\nu_T=3$, the quantum hall insulator is built on top of the quantum valley hall insulator at $\nu_T=2$.  

The IQHE state should have $C$ chiral edge modes which contribute a Hall conductance $G=C \frac{e^2}{h}$. For $\nu_T=3$, these additional $C$ chiral modes are robust, compared to the $2C$ helical modes from IQVHE effect. Also, the system may  form  domains with opposite valley polarization  which have opposite Chern number and Hall conductivity.  It may be possible to align the domains by cooling in  a magnetic field. We expect typical behavior of hysteresis loops of Hall resistance  for QAHE to be observed in a  magnetic field\cite{he2018topological}.

For both $\nu_T=1$ and $\nu_T=3$,  when the state is both spin  and valley polarized, there is a skyrmion defect in  the spin ferromagnetic order parameter. Following standard arguments, such a skyrmion  carries charge $Q_c=C$. Valley polarization breaks time reversal but preserves $U(1)_\upsilon$ symmetry. So there is no meron excitation associated with valley. For $C=2$, the skyrmion is a charge $2e$ boson while  for $C=3$ it is a charge $3e$ fermion.  Similar to quantum Hall systems, the skyrmion excitation is actually  the cheapest charged excitation for this insulating state in some parameter range (see Appendix.~\ref{appendix:skyrmion} for detailed calculation).  Thus for $C = 2$ these QAH insulating states at $\nu_T = 1,3$ have pair binding: we can regard the bosonic charge-$2e$ skyrmion as a Cooper pair. 
It is interesting to consider the effects of doping away from $\nu_T = 1$ or $\nu_T = 3$.  For small dopings the excess charge will be accommodated by introducing skyrmions.  A natural possibility, which is well established in $\nu=1$ quantum hall systems\cite{sondhi1993skyrmions}, is that the skyrmions form a lattice. It is interesting however to consider the alternate possibility that,  for $C=2$ systems, the bosonic skyrmions  condense into a liquid, instead of forming skyrmion lattice.  Due to the charge-$2e$ on the skyrmion, this is a  superconducting phase.  It is interesting to search for this kind of skyrmion superconductor close to $\nu_T=1$ or $\nu_T=3$ in BG/h-BN system or twisted BG/BG system. If such a superconductor is indeed found close to the ferromagnetic quantum hall insulator, the phase diagram may be quite interesting. There may be two energy scales: $T_c$,  the critical temperature of superconductor and $T^*$,  the energy scale for single electron gap. Below $T^*$ there are only bosonic excitations. $T_c$ should be proportional to the skyrmion density (thus doping level), while $T^*$ decreases with doping. At small doping, we expect $T_c << T^*$.  At intermediate temperatures  $T_c<T<T^*$, we may find a skyrmion metal with ``gapped"  fermionic excitation. Skyrmion condensation provides a mechanism to  get superconductivity starting from an insulator, instead of normal fermi liquid.

\section{Fractional Quantum Anomalous Hall Effect}
We now briefly consider the physics at fractional fillings $\nu_T$.  Based on some simple observations, we suggest situations  which are ripe for realization of a Fractional Quantum Anomalous Hall Effect (FQAHE), though a very clean sample and strong correlation may be necessary.

Upon electron doping the spin and valley polarized state at $\bar{\nu}_T=1$ or $\bar{\nu}_T=3$,  the next band to be filled is likely the one with the same spin and opposite valley  because of Hund's coupling. This also gives fractional filling of a Chern band with effective filling $\nu_{eff}=\nu_T- \bar{\nu}_T$. Therefore it is promising to search for valley polarized FQAHE states at these fillings.

 It has been established that there is a series of Abelian states at filling $\nu_{eff}=\frac{m}{2km |C|+1}$ with $k,m$ integers for flat Chern band with Chern number $C$\cite{wang2012fractional,liu2012fractional,sterdyniak2013series,moller2015fractional}. For example, if $k=1,m=-1$, we  can have an Abelian FQHE state at $\nu_{eff}=\frac{1}{2|C|-1}$.  For $k=1,m=1$, FQHE state happens at $\nu_{eff}=\frac{1}{2|C|+1}$. These states will have a quantized electrical  Hall conductivity $\sigma_{xy}^c=(\bar{\nu}_T -\nu_{eff})C$ where we also include the contribution from QAHE at filling $\bar{\nu}_T=1$ or $3$. The fully filled and partially filled bands reside in opposite valleys  which have opposite Chern numbers - thus their Hall conductivities subtract.

Other topological ordered states may also be possible in these nearly flat $\pm$ Chern bands, including those with  non-Abelian quasiparticles and fractional  topological insulators.

\section{Similarities with twisted bilayer graphene} 

When the bands have zero Chern number, it is possible to build a tight binding model by constructing Wannier orbitals for each valley in triangular lattice. A discussion of TLG/h-BN in this case was already given in Ref.~\onlinecite{po2018origin} and the resulting theory is a Hubbard model on triangular lattice with both spin and valley degrees of freedom at each site. In this paper, our focus is on the cases  with nonzero Chern number for each valley. For this case, it is clearly not possible to construct maximally localized Wannier orbital for each valley. This feature is also present - in a more subtle form - in the treatment in Ref. \onlinecite{po2018origin} of the small angle  twisted bilayer graphene system. In that case there are two bands per valley which have Dirac crossings. The obstruction to constructing maximally localized Wannier functions can be traced to the same sign chirality of the two Dirac points within a single valley. The systems with $\pm C$ Chern bands studied in this paper thus provide a conceptually simpler version where the Wannier obstruction of valley filtered bands  is ``obvious".  These systems lend further credence to this aspect of the theoretical treatment of Ref. \onlinecite{po2018origin} of the twisted bilayer graphene.

However, just like in Ref. \onlinecite{po2018origin} it should be possible to build Wannier orbitals for two valleys together because the total Chern number is $0$ for two valleys. The cost is that $U_v(1)$ valley symmetry can not be easily implemented in the resulting tight binding models. We leave the construction of this kind of tight binding model, using the methods introduced in Ref. \onlinecite{po2018origin},  to  future work. In this paper, we have instead tried to  understand the resulting states phenomenologically by working in momentum space directly.

Theoretically all these systems present a new experimental context in which the physics is controlled by correlation effects on bands with non-trivial topological aspects. This makes them likely different from ``ordinary" correlated solids which can be faithfully modeled through interacting lattice tight binding models with ordinary action of all physical symmetries.

\hfill

\section{Conclusion}
In conclusion, we propose graphene based moir\'e super-lattice systems to get nearly flat $\pm$ Chern bands with Chern number $|C|=2,3$.  Remarkably the Chern number can be controlled simply by a vertical electric field which further also tunes the bandwidth, and enhances interaction effects. This should enable not only a study of correlated $\pm$ Chern bands but also to cleanly experimentally isolate the effects of band Chern number from other effects in an interacting system. We proposed $\nu_T=2,4$ and $\nu_T=1,3$ insulating phases to be  Quantum Valley Hall insulator and Quantum Anomalous Hall insulator respectively in the limit of strong interaction. For other rational fractional fillings, we propose to search for FQAE states and FQVHE states. Realization of such exotic topological ordered states at zero field could provide a powerful platform to have more control over anyons, for example, to explore non-Abelian anyons bounded to lattice defects\cite{barkeshli2012topological}. For $C=\pm 2$ system, we also suggest the possibility of an exotic superconductor from condensation of charge $2e$ bosonic skyrmions close to filling $\nu_T=1$ or $\nu_T=3$. Finally we note conceptual similarities between the systems with $\pm C$ Chern bands studied here and the twisted bilayer graphene system.

\section{acknowledgement}
We thank  Valla Fatemi,  and Joe Checkelsky for many inspiring discussions.   TS is grateful to Adrian Po, Liujun Zou, and Ashvin Vishwanath for a previous collaboration which paved the way for the present paper.  Y-H.Z thanks Shaowen Chen for very helpful explanations of experimental details.
Y.H. Z, DM, and TS were supported by NSF grant
DMR-1608505, and partially through a Simons Investigator
Award from the Simons Foundation to Senthil Todadri. YC and PJH were primarily supported by the Gordon and Betty Moore Foundations EPiQS Initiative through grant GBMF4541 and the STC Center for Integrated Quantum Materials (NSF grant number DMR-1231319).

\bibliographystyle{apsrev4-1}
\bibliography{Chern}

\onecolumngrid
\appendix

\section{Methods for Modeling Moir\'e Mini Bands \label{appendix:methods}}
\subsection{Band Structure}
For the calculation of band structure, we follow the continuum model approach in Ref.~\onlinecite{jung2014ab} and Ref.~\onlinecite{bistritzer2011moire}. If the two layers have different lattice constants with $\xi=\frac{a_1-a_2}{a_2}$ or a twist angle $\theta$, there is a moir\'e super lattice with lattice constant $a_M\approx \frac{a}{\sqrt{\xi^2+\theta^2}}$. For the systems considered here, $a_M\sim 50-60 \sim 10-15$ nm.  In the limit $a_M>>a$,  it is appropriate to work in the continuum limit.  We ignore inter-valley scatterings and treat the two valleys separately. This is an excellent approximation for small twist angles. The two valleys are related by time reversal transformation. Therefore, we can do calculations for only one valley, for example, valley $+$.

We start from a continuum model of graphene for the valley $+$:
\begin{equation}
  H_0=\sum_{\mathbf k}c^\dagger_a(\mathbf k)h_{ab}(k)c_b(\mathbf k)
\end{equation}
where $a,b$ are collection of layer and orbital indexes.

Then moir\'e lattice gives super-lattice potentials: 
\begin{equation}
   H_M=\sum_{ab;\mathbf{k},\mathbf{G_j}}c_a^\dagger(\mathbf k)V_{ab}(\mathbf G_j)c_b(\mathbf{k+G_j})+h.c.
 \end{equation} 
 where $G_j$ is the moir\'e super-lattice reciprocal vector. We choose $\mathbf{G_1}=(0,\frac{4\pi}{\sqrt{3}a_M})$ and $\mathbf{G_2}=(-\frac{2\pi}{\sqrt{3}a_M},\frac{2\pi}{a_M})$ for the moir\'e Brillouin zone (MBZ) of the moir\'e super-lattice. The form of the $H_M$ is model dependent and details are shown in Section.~\ref{section:BG/hBN} and Section.~\ref{appendix:graphene_graphene}.
 
 The moir\'e potential reconstructs the original band  into  a small Moir\'e Brillouin Zone(MBZ) which is a hexagon with size $|K|=\frac{4\pi}{3 a_M}$. The calculation of band structure is very similar to the classic example of  free electron approximation in elementary solid state textbooks\cite{ashcroft2005solid}. For each momentum $\mathbf{k}$ in the MBZ, we build a large matrix $H(k)$ whose bases include states for $M*M$ momentum points: $\mathbf k+m \mathbf G_1 +n \mathbf G_2$, $m,n=-M/2,...,M/2$. Typically we use $M=11$ for calculation of band structure and $M=7$ for calculation of Berry curvature. Diagonalization of $H(k)$ gives both the energy dispersion and the Bloch wave function $\ket{\mu(\mathbf k)}$.

 We focus on the conduction and valence bands close to the neutrality point.  In contrast to twisted bilayer graphene ({\em i.e} monolayer on monolayer), here there is no $C_6$ (and hence inversion symmetry) rotation symmetry even approximately. Hence we do not expect any generic protected crossings  between the valence and conduction bands.

  \subsection{Numerical calculation of Berry Curvature and Chern number\label{section:Appendix_Berry}}

  In addition to the band structure, we can also easily calculate  the Berry curvature and Chern number of the valence band.

Our calculation of Berry curvature and Chern number follows Ref.~\onlinecite{fukui2005chern}.
We discretize the Moir\'e Brillouin Zone(MBZ) with a $N_x \times N_y$ grid and then we have a lattice in momentum space. For each momentum $\mathbf k$ in MBZ, we diagonalize the large matrix $H(\mathbf k)$ and extract the eigenvector  $\mu(\mathbf k)$ corresponding to  the valence band. We calculate the following two complex quantities:
\begin{equation}
  U_{\hat{x}}(\mathbf k)=\braket{ \mu(\mathbf k +\delta k_x \hat k_x)|\mu(\mathbf k)}
\end{equation}
and
\begin{equation}
  U_{\hat{y}}(\mathbf k)=\braket{ \mu(\mathbf k +\delta k_y \hat k_y)|\mu(\mathbf k)}
\end{equation}
where $\hat k_x$ and $\hat k_y$ are unit vectors in momentum space in $k_x$ and $k_y$ directions. $\delta k_x$ and $\delta k_y$ are lattice constants in $k_x$ and $k_y$ directions for our momentum lattice. The phases of $U_{\hat x}$ and $U_{\hat y}$ are Berry connections on the links of our momentum space lattice. They are not gauge invariant under a gauge transformation $\mu(\mathbf k) \rightarrow \mu(\mathbf k)e^{i\theta(\mathbf k)}$. However, we can calculate the Wilson loop for one plaquette at $\mathbf k$:
\begin{equation}
  W(\mathbf k)=U_{\hat x}(\mathbf k)U_{\hat y}(\mathbf k+\delta k_x \hat k_x)U^*_x(\mathbf k +\delta k_y \hat k_y)U^*_y(\mathbf k)
\end{equation}

$W(\mathbf k)$ is gauge invariant and Berry curvature at $\mathbf k$ can be approximated as :
\begin{equation}
  \mathbf{B}(\mathbf k)=\frac{\arg W(\mathbf k)}{\delta k_x \delta k_y}
\end{equation}

We emphasize that we need to guarantee that $|\arg W(\mathbf k)|<\pi$ for any $\mathbf k$. This condition is crucial to avoid ambiguity in the calculation. So long as Berry curvature has no singularity, we can always fulfill this condition by reducing the size of the plaquette. The error of this approximation can be reduced by increasing $N_x$ and $N_y$, which decreases $\delta k_x$ and $\delta k_y$. The Chern number is calculated by:
\begin{equation}
  C=\frac{1}{2\pi}\sum_{\mathbf {k}}\mathbf{B}(\mathbf k)\delta k_x \delta k_y
\end{equation}

 To get accurate results in our continuum model, we also need $M\rightarrow \infty$. Therefore, the calculation of Chern number is exact at the limit $N_x,N_y, M \rightarrow \infty$.  Fortunately, we found that $M=3-5$, $N_x,N_y=40$ is enough to get an accurate Chern number.

\section{Models for Bilayer Graphene on Boron Nitride \label{section:BG/hBN}}

We give some details of our continuum models for calculation of BG/h-BN or TG/h-BN systems, following Ref.~\onlinecite{jung2014ab}.

\subsection{Moir\'e Hamiltonians}

 We label the microscopic real space electron operators as $c_{A_1}$, $c_{B_1}$, $c_{A_2}$, $c_{B_2}$. We assume $B_1$ is on top of $A_2$.   

For convenience, we label $\psi_a=\left(\begin{array}{c} c_{A_a}\\ c_{B_a} \end{array}\right)$ where $a=1,2$ is layer index.

The Hamiltonian is
\begin{equation}
	H=H_0+H_M
\end{equation}
where
\begin{equation}
	H_0=t\sum_a\sum_k \psi^\dagger_a(k)(k_x \sigma_x+k_y \sigma_y)\psi_a(k)+\gamma_1\left(\sum_k \psi_1^\dagger(k) \sigma_{-}\psi_2(k)+h.c.\right)+t_3\left(\sum_k (k_x+ik_y)\psi_1^\dagger(k) \sigma_+ \psi_2(k)+h.c.\right)
\end{equation}
where $t=2676$meV and $\gamma_1=340$meV. $t_3=0.051 t$ is the trigonal warping term.

The moir\'e super-lattice potential acts only on the first layer, which is closest to h-BN.
\begin{equation}
	H_M=\sum_{\mathbf{G}}\psi_1^\dagger(\mathbf{k}+\mathbf{G}) H_V(\mathbf{G})\psi_1(\mathbf k)+U \psi_1^\dagger(\mathbf k) \psi_1(\mathbf k)
\end{equation}
 $U$ is the energy difference between top layer and bottom layer, which is controlled by perpendicular electric field.

 $H_V(\mathbf G)$ can be expanded in terms of $I,\sigma_x,\sigma_y,\sigma_z$
 \begin{equation}
 	H_V(\mathbf G)=H_0(\mathbf G)+H_z(\mathbf G)\sigma_z+ Re(H_{AB}(\mathbf G))\sigma_x+Im(H_{AB}(\mathbf G))\sigma_y
 \end{equation}

In the following we list parameters for $H_V(\mathbf{G})$. They are fit from the ab initio calculation in Ref.~\onlinecite{jung2014ab}.

\begin{align}
&H_{a}(\mathbf{G_1})=H_{a}(\mathbf{G_3})=H_{a}(\mathbf{G_5})=C_{a}e^{i\varphi_a} \notag\\
&H_{a}(\mathbf{G_2})=H_{a}(\mathbf{G_4})=H_{a}(\mathbf{G_6})=C_{a}e^{-i\varphi_a} 
\label{eq:Gbn-diagonal_terms}
\end{align}
where $a=0,z$. We have $C_0=-10.13$meV, $\varphi_0=86.53^\circ$ and $C_z=-9.01$meV, $\varphi_z=8.43^\circ$.

For $H_{AB}$, we have
\begin{align}
&H_{AB}(\mathbf{G_1})=H^*_{AB}(\mathbf{G_4})=C_{AB}e^{i(\frac{2\pi}{3}-\varphi_{AB})}\notag\\
&H_{AB}(\mathbf{G_3})=H^*_{AB}(\mathbf{G_2})=C_{AB}e^{-i\varphi_{AB}}\notag\\
&H_{AB}(\mathbf{G_5})=H^*_{AB}(\mathbf{G_6})=C_{AB}e^{i(-\frac{2\pi}{3}-\varphi_{AB})}
\label{eq:Gbn-AB_terms}
\end{align}
where $C_{AB}=11.34$meV and $\varphi_{AB}=19.60^\circ$.

All possible $\mathbf{G}$ can be written as
\begin{equation}
  \mathbf{G}=m \mathbf{G_1}+ n\mathbf{G_2}
\end{equation}
where $\mathbf{G_1}=(0,\frac{4\pi}{\sqrt{3}a_M})$ and $\mathbf{G_2}=(-\frac{2\pi}{a_M},\frac{2\pi}{\sqrt{3}a_M})$.

ABC Trilayer graphene on Boron Nitride can be modeled similarly. Because we can assume that only the layer closest to the h-BN layer is influenced, we can use the same parameters as above.

\subsection{Gate Tunable Band Dispersions}

\begin{figure}[ht]
\centering
  \begin{subfigure}[b]{0.45\textwidth}
    \includegraphics[width=\textwidth]{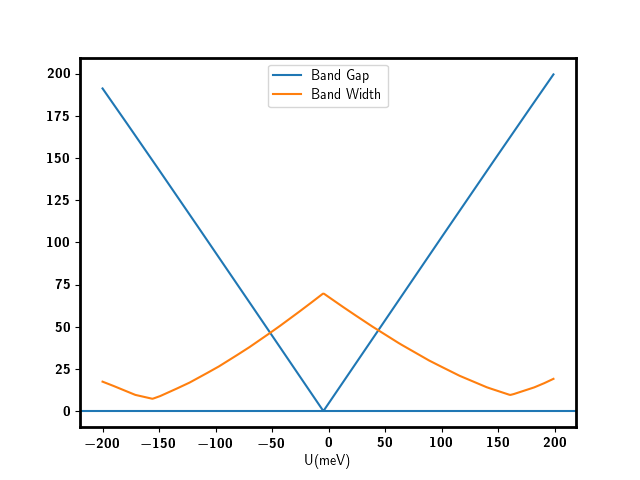}
    \caption{BG/h-BN}
  \end{subfigure}
  \begin{subfigure}[b]{0.45\textwidth}
    \includegraphics[width=\textwidth]{image/TG_gap_U.png}
    \caption{TG/h-BN}
  \end{subfigure}
  
  \caption{Band width and band gap (in units of meV) with applied voltage difference $U$.}
  \label{fig:gate_driven_band_width}
\end{figure}

Qualitatively, different direction of applied vertical field does not change band dispersion, as is obvious in Fig.~\ref{fig:gate_driven_band_width}.  Increasing $|U|$ can reduce  the bandwidth and increase the band gap between the conduction and the valence bands.

\subsection{Chern number of BG/h-BN and TG/h-BN systems \label{Appendix:chern_number_bg_hbn}}
\begin{figure}[ht]
\centering
  \begin{subfigure}[b]{0.45\textwidth}
    \includegraphics[width=\textwidth]{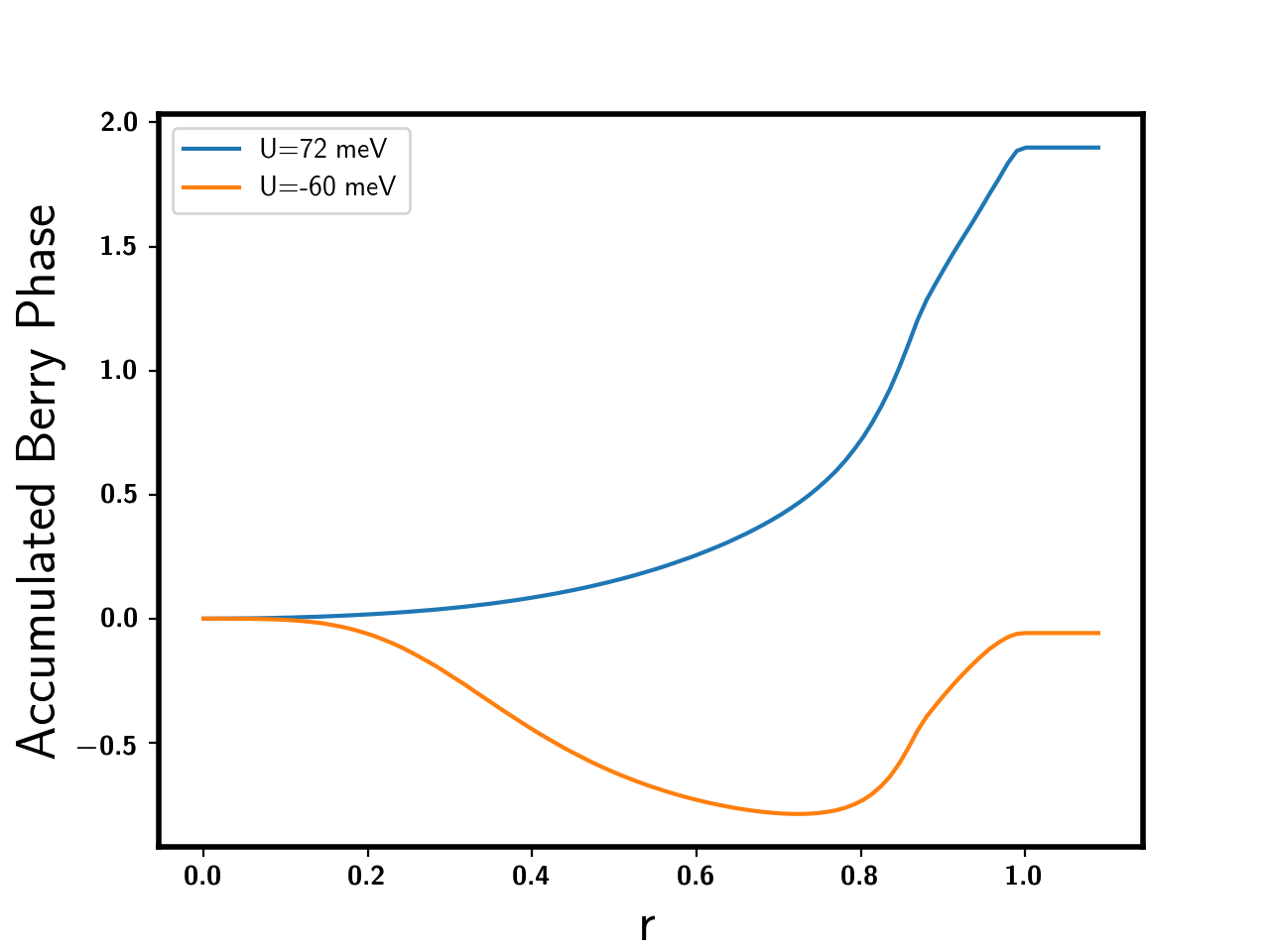}
    \caption{BG/h-BN}
  \end{subfigure}
  \begin{subfigure}[b]{0.45\textwidth}
    \includegraphics[width=\textwidth]{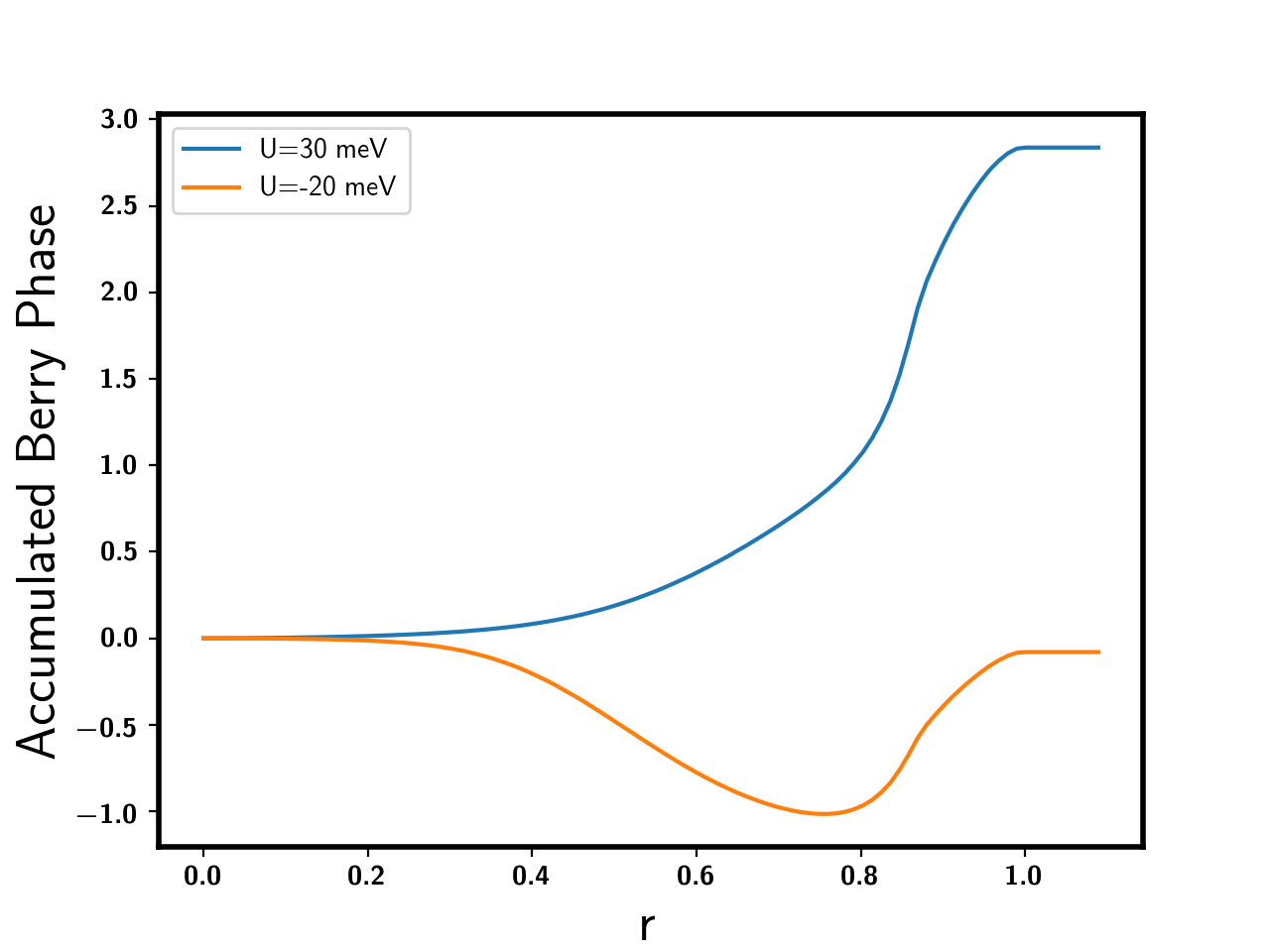}
    \caption{TG/h-BN}
  \end{subfigure}
  \caption{Integration of Berry curvature in region within MBZ  with distance smaller than $r$ to $\Gamma$ point. $r$ has been normalized by $|\mathbf K_M|=\frac{4\pi}{3 a_M}$. The value at $r=1$ is the Chern number.}
  \label{fig:Berry_phase}
\end{figure}

 The Chern number for valence band of valley $+$ jumps by $\Delta C=n$ when changing $U$ from negative to positive. To extract the value of Chern number of each side, we perform  numerical calculation as explained in Appendix.~\ref{section:Appendix_Berry}.  As shown in Fig.~\ref{fig:Berry_phase}, Chern number $(C(U<0),C(U>0))$ is $(0,2)$ for BG/h-BN and $(0,3)$ for TG/h-BN.

We also want to discuss the stability of Chern number at the band structure level, i.e. whether it changes for other parameters of moir\'e super-lattice potential terms. 

For this purpose, we can use the two band continuum model  model in Eq.~\ref{eq:h0} for BG and TG before adding moir\'e potentials.

The main effect from  the moir\'e pattern is a moir\'e super-lattice potential for the first graphene layer. It can always be written as:

\begin{equation}
  H_M=\sum_{\mathbf {G}} c^\dagger_{a1}H_{ab}(\mathbf G)c_{b1}
  \label{eq:moir\'e_layer1}
\end{equation}
where $a,b=A,B$ is sublattice index for first graphene layer.

At low energy for  BG, only $A_1$ and $B_2$ need to be kept.  Projecting to these lowest bands $A_1$ and $B_2$, only $H_{AA}(\mathbf G)$ in Eq.~\ref{eq:moir\'e_layer1} is kept for layer 1. The low energy mode can be expressed as a  spinor $\psi=\left(\begin{array}{c}c_t\\c_b\end{array}\right)$ with $c_t=c_{A_1}$ and $c_t=c_{B_2}$. For ABC stacked TG, similarly $(B_1,A_2)$ and $(B_2,A_3)$ are dimerized. We only keep $c_t=c_{A_1}$ and $c_b=c_{B_3}$. $h-BN$ is a strong insulator and therefore all of electronic degrees of freedom come from the graphene. We consider the systems for which only h-BN layer on top of the graphene is aligned. As a result, the superlattice potential only acts on the the first layer of BG or TG.  For both BG/h-BN and TG/h-BN,  the only important term from the moir\'e super-lattice is:

\begin{equation}
  H_M=\sum_{\mathbf {G_j}}\sum_{\mathbf k} \psi^\dagger(\mathbf k+\mathbf G_j)\left( \begin{array}{cc}V_0e^{i\theta_j}&0\\0&0\end{array}\right)\psi(\mathbf k)+h.c.
  \label{eq:simple_moir\'e}
\end{equation}

In summary, for both BG and TG, essential properties of band structure are well described by Eq.~\ref{eq:h0} and Eq.~\ref{eq:simple_moir\'e}. Only two parameters $V_0$ and $\theta$ are important. In this system $C_2$ symmetry is broken and the only point group symmetry remaining is $C_3$ rotation.

Then with similar methods as in previous sections, we calculate Berry curvature by keeping $M^2$ momentum points and $2M^2$ bands. With this simple model, we study the full phase diagram as a function of $V_0$ and $\theta_0$. The result is that Chern number is quite stable. Chern number in one direction of perpendicular electric field($E_z<0$) is robustly $C=2$ for BG/h-BN and $C=3$ for TG/h-BN.

\section{Twisted Graphene on Graphene Systems\label{appendix:graphene_graphene}}

In this section we discuss the possibility of nearly flat Chern bands in twisted graphene+graphene systems  close to the magic angle($\theta_M=1.08^\circ$ for BG/BG and TG/TG; $\theta_M=1.24^\circ$ for TG/TG). Each graphene layer is either AB stacked bilayer graphene or ABC stacked trilayer graphene. We will study the following three different systems: BG/BG,TG/TG and TG/BG. Typically there are two different patterns of stacking, depending on whether two graphenes have the same chirality or not.  We define the stacking pattern with the same chirality as type I and the other one with opposite chirality as type II. Taking BG/BG as an example, we call AB on AB as type I and AB on BA as type II.  In the following we give a short summary of our results for BG/BG system with type I stacking. The other two systems have similar magic angles. For each specific valley, we place the original quadratic band touching point of one BG at $K$ point of  the MBZ, and place the band touching point of the other BG at $K'$ of the MBZ. We use the same model for moir\'e hopping terms between the middle two layers as in Ref.~\onlinecite{bistritzer2011moire}. There is only one parameter $t_M$ in this model. We use $t_M=110$ meV which is fit from ab initio numerics\cite{jung2014ab}.

\begin{figure}[ht]
\centering
\begin{subfigure}[b]{0.45\textwidth}
    \includegraphics[width=\textwidth]{image/bg_bg_conduction_band_width.png}
    \caption{Bandwidth of conduction band with twist angle.}
	\label{fig:magic_BG_BG}
  \end{subfigure}
  \begin{subfigure}[b]{0.45\textwidth}
    \includegraphics[width=\textwidth]{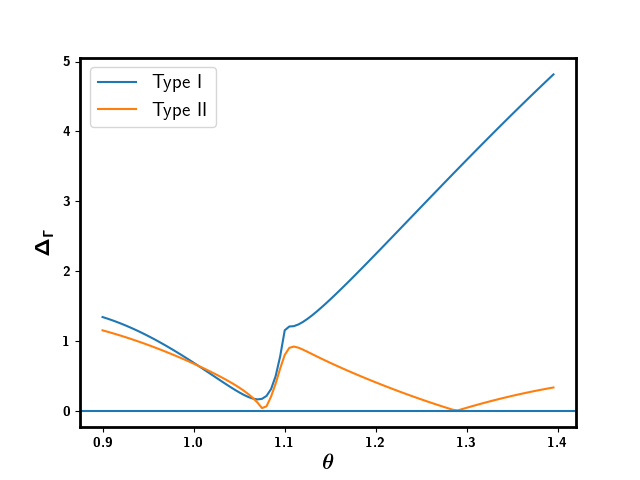}
	\caption{The gap $\Delta_\Gamma$(in unit of meV) between conduction band and the band above at $\Gamma$ point. }
	\label{fig:gamma_gap_BG_BG}
  \end{subfigure}
  \caption{BG/BG system with both type I and type II stackings. At magic angle $\theta \approx 1.08^\circ$, band width reduces to almost zero. Meanwhile the hybridization gap between conduction band and the band above also goes to zero. Roughly $U_c(\theta)\sim \Delta_\Gamma(\theta)$. AB on BA stacking has very different $U_c(\theta)$ when $\theta>\theta_M$.  $t_M=110$ meV in these calculations.}
  \label{fig:magic_angle}
\end{figure}

\begin{figure}[ht]
\centering
  \begin{subfigure}[b]{0.3\textwidth}
    \includegraphics[width=\textwidth]{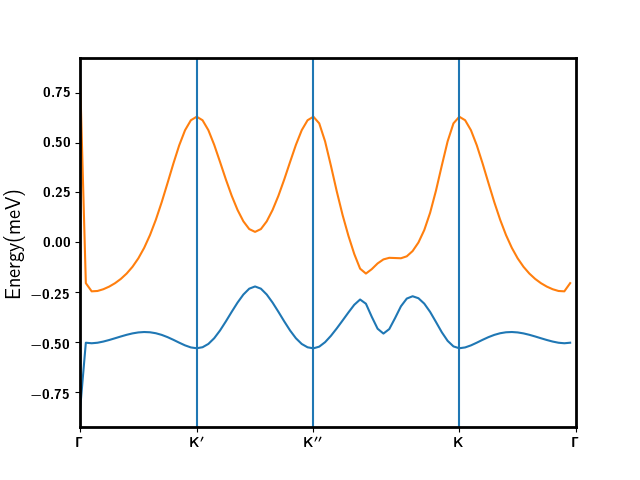}
    \caption{ $U=0$ meV}
  \end{subfigure}
  \begin{subfigure}[b]{0.3\textwidth}
    \includegraphics[width=\textwidth]{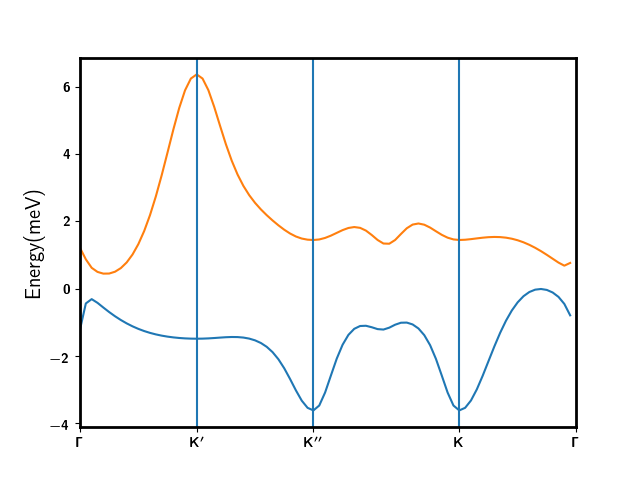}
    \caption{ $U=10$ meV}
  \end{subfigure}
  
  \begin{subfigure}[b]{0.3\textwidth}
    \includegraphics[width=\textwidth]{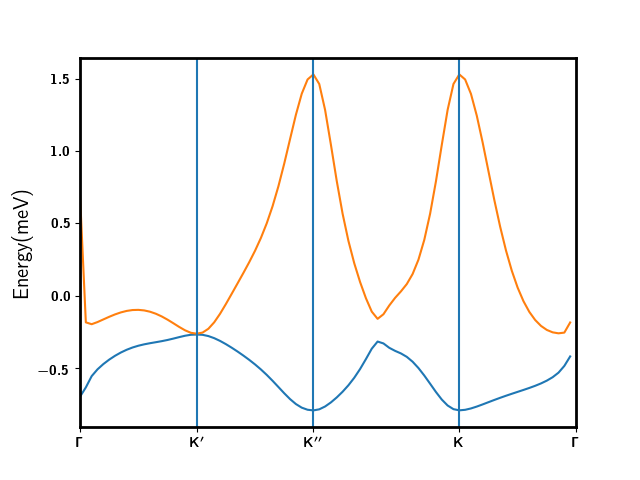}
    \caption{ $U=-1.8$ meV}
  \end{subfigure}
   \begin{subfigure}[b]{0.3\textwidth}
    \includegraphics[width=\textwidth]{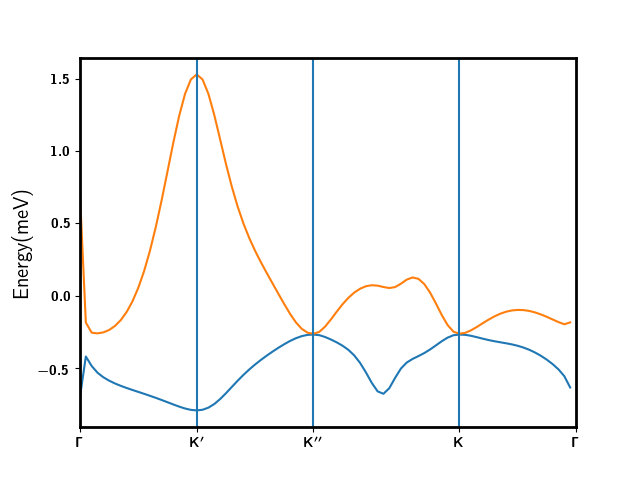}
    \caption{$U=1.8$ meV}
  \end{subfigure}
  \caption{Band structures for type I BG/BG system. Twist angle is $\theta=1.08^\circ$. $K'=(0,\frac{4\pi}{3a_M})$. $K''=(\frac{2\pi}{\sqrt{3}a_M},\frac{2\pi}{3a_M})$ and $K=(0,-\frac{4\pi}{3a_M})$ are equivalent. At $U=\pm 1.8$ meV, mass term of one cone at $K$ or $K'$ changes sign.}
  \label{fig:bg_bg_dispersion}
\end{figure}

First, at $U=0$, we found that the original band touching point is gapped out. The gap is quite small: $\Delta \lessapprox 1.5$ meV. Similar to the twisted monolayer graphene on monolayer graphene system \cite{bistritzer2011moire}, there is magic angle where bandwidth goes to below $1$ meV, as shown in Fig.~\ref{fig:magic_BG_BG}.  For $t_M=110$ meV, magic angle is at $\theta\approx 1.08^\circ$.   For type I BG/BG, we show the band structure at $\theta=1.08^\circ$ for several perpendicular $U$ in Fig.~\ref{fig:bg_bg_dispersion}. At $U=0$, two cones at $K$ and $K'$ are gapped with different sign of mass. After we add positive $U=1.8$ meV, the cone  at $K$ with negative mass closes its gap and changes to positive mass. Similarly if we add negative  $U=-1.8$ meV, the cone at $K'$ with positive mass closes its gap and change mass to negative.

Let us focus on the region far away from  the magic angle. If we neglect the moir\'e hopping term $t_M$, there is a band crossing described by Eq.~\ref{eq:h0} with chirality $n_1$ at $K$ point and there is another band crossing with chirality $n_2$ at $K'$ point of MBZ. For BG/BG with AB on AB stacking pattern, $n_1=n_2=2$. When we change $U$ from negative to positive, both band crossings change mass from negative to positive. The sign change of mass can be seen in Fig.~\ref{fig:bg_bg_dispersion}.

\begin{figure}[ht]
\centering
  \begin{subfigure}[b]{0.3\textwidth}
    \includegraphics[width=\textwidth]{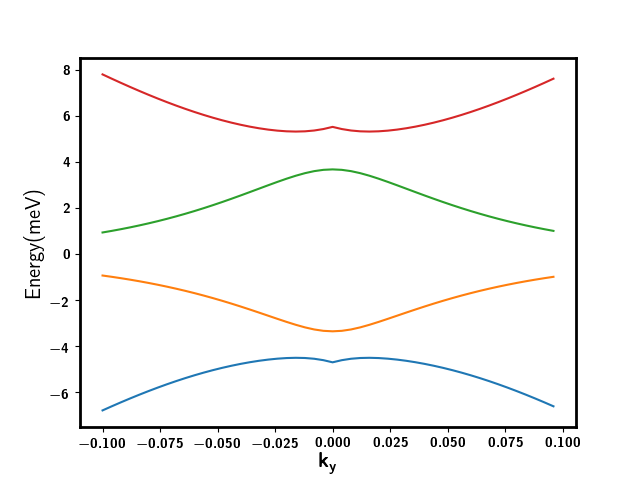}
    \caption{ $U=0$ meV}
  \end{subfigure}
  \begin{subfigure}[b]{0.3\textwidth}
    \includegraphics[width=\textwidth]{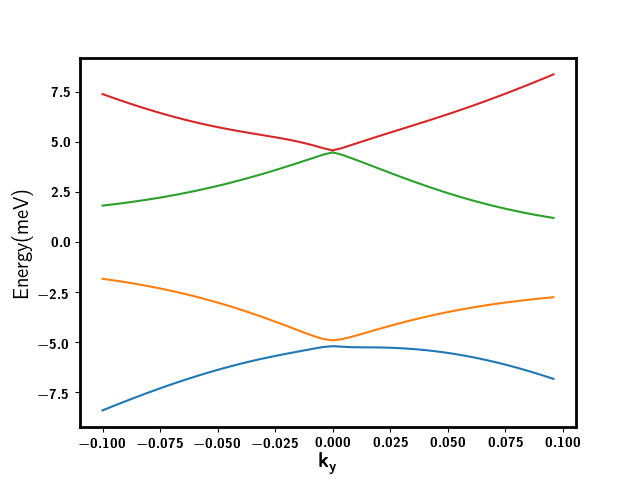}
    \caption{ $U=12$ meV}
  \end{subfigure}
  \begin{subfigure}[b]{0.3\textwidth}
    \includegraphics[width=\textwidth]{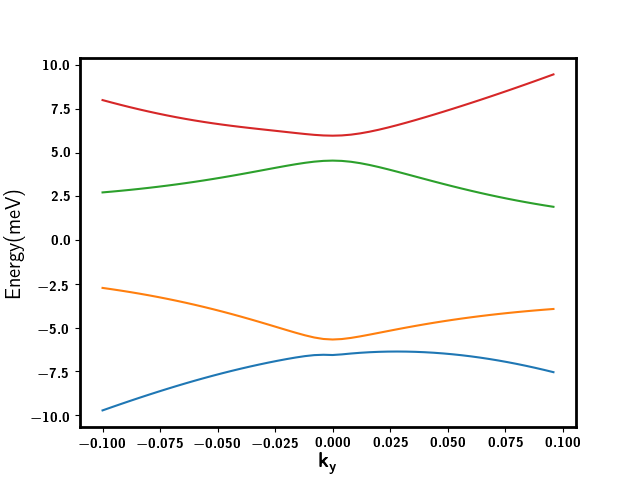}
    \caption{ $U=20$ meV}
  \end{subfigure}

  \caption{Dispersion around $\Gamma$ point for type I BG/BG system. Twist angle is $\theta=1.17^\circ$. Gap closing is obvious at $U_c(\theta)=12$ meV.}
  \label{fig:bg_bg_gamma_dispersion}
\end{figure}

There is another feature associated with the magic angle. For each twist angle $\theta$, there is a critical $U_c(\theta)$ through which the hybridization gap between conduction band and the band above closes. At magic angle $U_c(\theta)$ goes to zero. To see $U_c(\theta)$, we plot the dispersion around $\Gamma$ point for  type I BG/BG system at twist angle $\theta=1.17^\circ$ in Fig.~\ref{fig:bg_bg_gamma_dispersion}. Clearly at $U_c=12$ meV the hybridization gap between conduction band and the band above closes to a Dirac cone and reopens at $\Gamma$ point.

We have shown results of Chern numbers for BG/BG system in the main text. Here we list some results for TG/TG and TG/BG systems. We focus on type I stacking for these systems. For TG/TG system, there is a Mirror reflection symmetry $M_x$ which acts  in the same way as BG/BG system.

\begin{figure}[ht]
\centering
  \begin{subfigure}[b]{0.45\textwidth}
    \includegraphics[width=\textwidth]{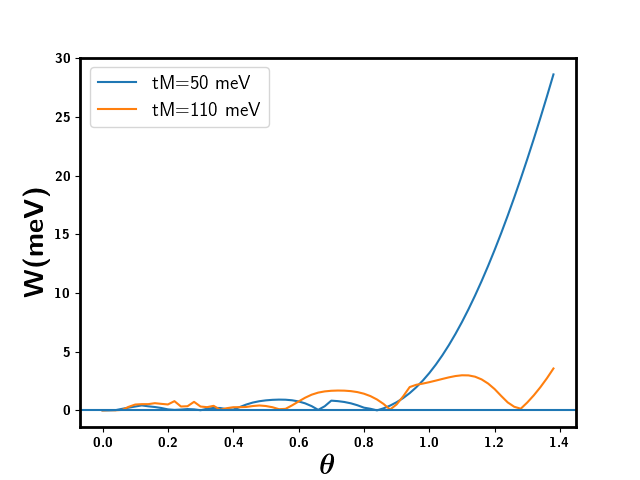}
    \caption{TG/TG}
  \end{subfigure}
  \begin{subfigure}[b]{0.45\textwidth}
    \includegraphics[width=\textwidth]{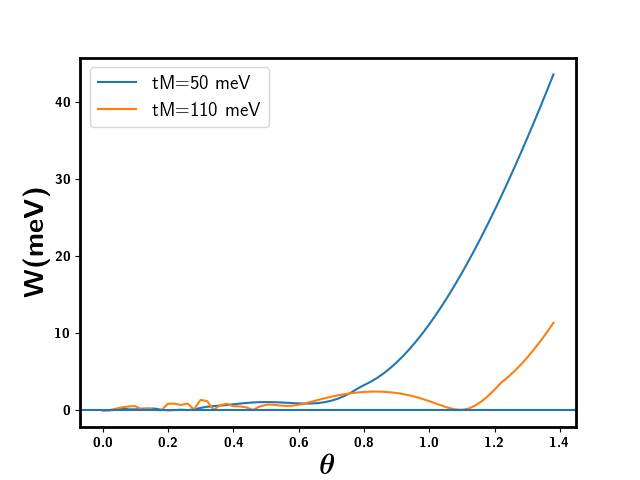}
    \caption{TG/BG}
  \end{subfigure}

  \caption{Magic angle for type I TG/TG and TG/BG systems. $t_M$ are hopping parameter for modeling moir\'e hopping term. Real system corresponds to $t_M=110$ meV.}
  \label{fig:append_magic_angle}
\end{figure}

We show the bandwidth for type I TG/TG and TG/BG system in Fig.~\ref{fig:append_magic_angle}. Clearly the first magical angle is at $\theta_M \approx 1.08^\circ$ for TG/BG while it is  at $\theta_M \approx 1.24^\circ$ for TG/TG.

\begin{figure}[ht]
\centering
	\begin{subfigure}[b]{0.3\textwidth}
    \includegraphics[width=\textwidth]{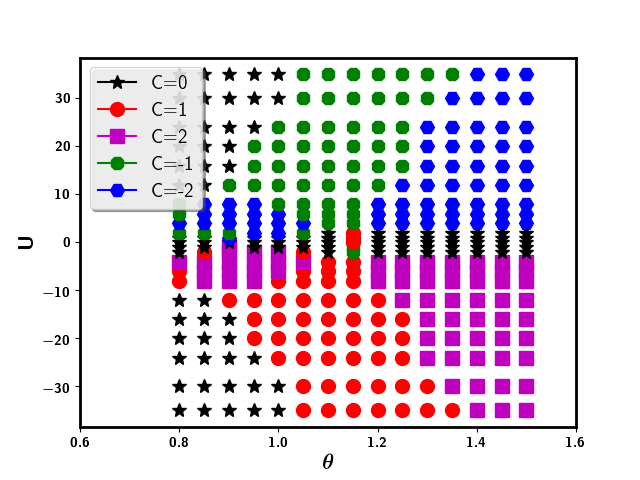}
    \caption{Type I BG/BG: conduction band}
  \end{subfigure}
  \begin{subfigure}[b]{0.3\textwidth}
    \includegraphics[width=\textwidth]{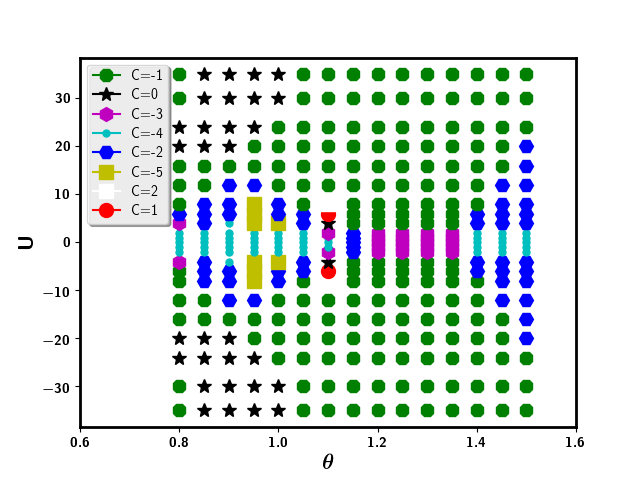}
    \caption{Type II BG/BG:conduction band}
  \end{subfigure}
	\begin{subfigure}[b]{0.3\textwidth}
    \includegraphics[width=\textwidth]{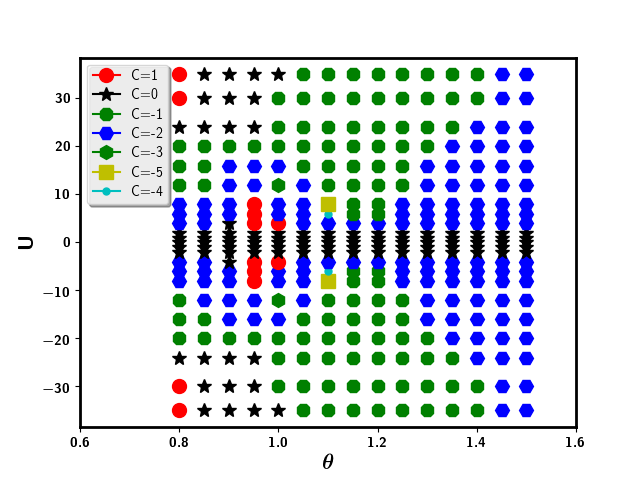}
    \caption{Type II BG/BG:valence band}
  \end{subfigure}

  \begin{subfigure}[b]{0.3\textwidth}
    \includegraphics[width=\textwidth]{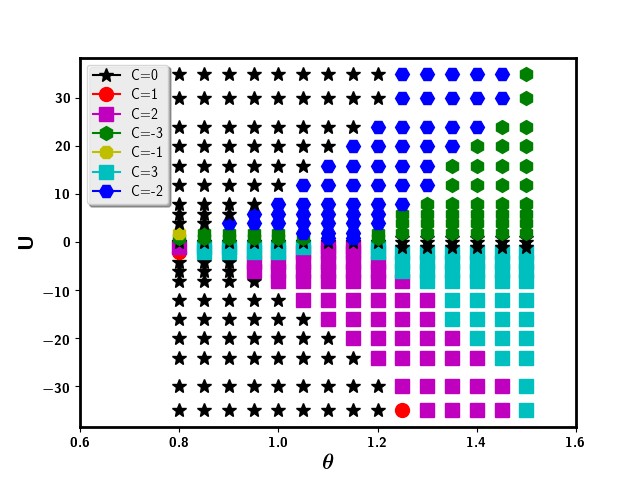}
    \caption{Type I TG/TG: conduction band}
  \end{subfigure}
  \begin{subfigure}[b]{0.3\textwidth}
    \includegraphics[width=\textwidth]{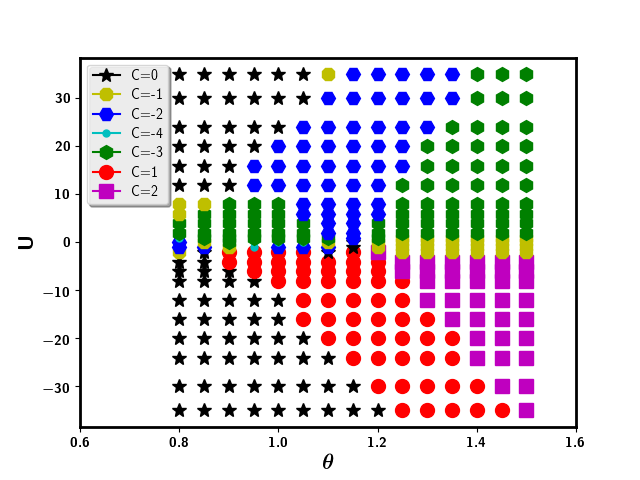}
    \caption{Type I TG/BG: conduction band}
  \end{subfigure}
  \begin{subfigure}[b]{0.3\textwidth}
    \includegraphics[width=\textwidth]{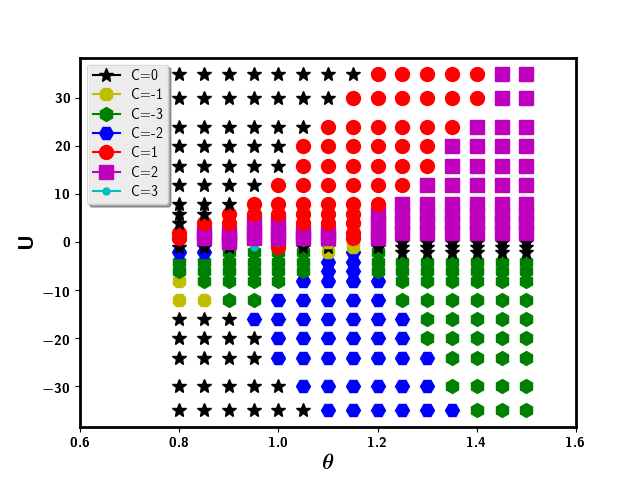}
    \caption{Type I TG/BG: valence band}
  \end{subfigure}

  \caption{Chern number $C$ for valley $+$ of twisted graphene on graphene systems.}
  \label{fig:append_abs_C}
\end{figure}

We also lists Chern number $C$ of valley $+$  for  BG/BG systems with both stacking patterns and type I TG/TG and TG/BG systems in Fig.~\ref{fig:append_abs_C}.  For type I BG/BG and TG/TG system, there is a reflection symmetry  which requires that $C(U)=-C(-U)$ and $C(U=0)=0$.

Below we give a more detailed description of models and symmetries for twisted graphene/graphene systems using BG/BG as an example.

\subsection{Model and Symmetry of BG/BG system}
We first show details of our models  for the following two different stacking patterns of BG/BG system: AB on AB (type I), AB on BA (type II).
\subsubsection{Type I BG/BG}
Here we perform a numerical simulation of continuum models. Again we focus only on one valley $+$; valley $-$ is just its time reversal partner. For top  and bottom BG, we define a four component spinor $\psi_a=\left(\begin{array}{c}c^a_{A_1}\\c^a_{B_1}\\c^a_{A_2}\\c^a_{B_2}\end{array}\right)$ where $a=t,b$ labels top BG or bottom BG layer. 

For valley $+$, $k^t$ of $\psi_t(k^t)$ is defined relative to the  band touching point of the original large Brillouin Zone of top layer and similarly $k^b$ of $\psi_b(k^b)$  is defined relative to the band touching point of bottom layer. In our set up, $k^t=0$ is at $K=(-\frac{2\pi}{\sqrt{3}a_M},-\frac{2\pi}{3})$ point of MBZ while $k^b=0$ is at $K'=(-\frac{2\pi}{\sqrt{3}a_M},\frac{2\pi}{3})$ point of MBZ. We use $k$ instead of $k^t$ or $k^b$ in the following just for simplicity.

The Hamiltonian of valley $+$ for twisted BG/BG system is:
\begin{align}
  H&=\sum_{a,\mathbf k} \psi_a^\dagger(\mathbf k)H^0_a(\mathbf k)\psi_a(\mathbf k)\notag\\
  &+\sum_{\mathbf k,\mathbf {Q_j}}\left( \psi^\dagger_t(\mathbf k) T_j\psi_b(\mathbf{k+Q_j})+h.c.\right)
\end{align}

where 
\begin{align}
&H^0_a(\mathbf k)\notag\\
&=\left(
\begin{array}{cccc}
U^a_1 &\upsilon(k_x-i k_y) &0 & t_3(k_x+i k_y)\\
\upsilon(k_x+i k_y)&U^a_1 &\gamma_1 &0\\
0&\gamma_1 &U^a_2&\upsilon(k_x-i k_y)\\
t_3(k_x-ik_y)&0&\upsilon(k_x+ik_y)&U^a_2
\end{array}
\right)
\end{align}
Here $U^a_1$ and $U^a_2$ means the energy of first  and second layers of top or bottom BG. This term comes from the applied voltage difference. $\gamma_1$ is the vertical hopping term between $B_1$ and $A_2$. $t_3$ is the trigonal warping term.

Moir\'e hopping matrix $T_j$ only hybridize $(c^t_{A_2},c^t_{B_2})$ with $(c^b_{A_1},c^b_{B_1})$. It is
\begin{equation}
  T_j=\left(
\begin{array}{cccc}
0 &0 &0 & 0\\
0&0 &0 & 0\\
t_M&t_M e^{-ij\varphi} &0&0\\
t_M e^{ij\varphi}&t_M&0&0
\end{array}
\right)
\label{eq:tM_AB_AB}
\end{equation}
where $\varphi=\frac{2\pi}{3}$ and $j=0,1,2$. 

One can check that this Hamiltonian is invariant under  $C_3$ rotation with $\mathbf k\rightarrow C_3 \mathbf k$, and
\begin{align}
c^t_{A_1}(\mathbf k) &\rightarrow c^t_{A_1}(C_3 \mathbf{k}) e^{i \varphi} \notag\\
c^t_{B_1}(\mathbf k) &\rightarrow c^t_{B_1}(C_3 \mathbf{k}) e^{-i \varphi} \notag\\
c^t_{A_2}(\mathbf k) &\rightarrow c^t_{A_2}(C_3 \mathbf{k}) e^{-i \varphi} \notag\\
c^t_{B_2}(\mathbf k) &\rightarrow c^t_{B_2}(C_3 \mathbf{k})  \notag\\
c^b_{A_1}(\mathbf k) &\rightarrow c^b_{A_1}(C_3 \mathbf{k}) e^{-i \varphi} \notag\\
c^b_{B_1}(\mathbf k) &\rightarrow c^b_{B_1}(C_3 \mathbf{k}) \notag\\
c^b_{A_2}(\mathbf k) &\rightarrow c^b_{A_2}(C_3 \mathbf{k})  \notag\\
c^b_{B_2}(\mathbf k) &\rightarrow c^b_{B_2}(C_3 \mathbf{k}) e^{ i \varphi} \notag\\
\end{align}

In addition to the $C_3$ symmetry, we have a mirror reflection symmetry $M_x$ which exchanges top and bottom graphene layers  without flipping the valley index.

The action of $M_x$ is:
\begin{equation}
\begin{array}{c}
c_{A_1}^t(k)\rightarrow c_{B_2}^b(M_x k)\\
c_{B_1}^t(k)\rightarrow c_{A_2}^b(M_x k)\\
c_{A_2}^t(k)\rightarrow c_{B_1}^b(M_x k)\\
c_{B_2}^t(k)\rightarrow c_{A_1}^b(M_x k),
\end{array}
\end{equation}
where $M_x (k^t_x, k^t_y)= (k^b_x, -k^b_y)$. It is easy to check that $H_0$ is invariant under the above transformation. For the terms in $T_j$, we have,
\begin{equation}
\begin{array}{c}
c_{A_2}^{t\dag}(k)c_{A_1}^b(k+Q_j)\rightarrow c_{B_1}^{b\dag}(M_x k)c_{B_2}^t(M_x(k+Q_j))\\
c_{B_2}^{t\dag}(k)c_{B_1}^b(k+Q_j)\rightarrow c_{A_1}^{b\dag}(M_x k)c_{A_2}^t(M_x(k+Q_j))\\
c_{A_2}^{t\dag}(k)c_{B_1}^b(k+Q_j)e^{-ij\varphi}\rightarrow c_{B_1}^{b\dag}(M_x k)c_{A_2}^t(M_x(k+Q_j))e^{-ij\varphi}\\
c_{B_2}^{t\dag}(k)c_{A_1}^b(k+Q_j)e^{ij\varphi}\rightarrow c_{A_1}^{b\dag}(M_x k)c_{B_2}^t(M_x(k+Q_j))e^{ij\varphi},
\end{array}
\end{equation}
Note that $M_x Q_{0,1,2}\rightarrow -Q_{0,2,1}$, so the Hamiltonian $H(k)\rightarrow H(M_x k)$.

In experiments  an energy difference $U$ between top layer and bottom layer can be induced by a perpendicular electric field. We simply model this energy difference by adding an on-site energy $U^t_1=U$, $U^t_2=\frac{2 U}{3}$, $U^b_1=\frac{U}{3}$ and $U^b_2=0$. Basically we assume voltage increase $V_{12}=V_{23}=V_{34}=\frac{U}{3 e}$, where $1,2,3,4$ labels the four  layers from top to bottom.

It is easy to see that $M_x$ maps $U$ to $-U$ (up to an constant energy shift). Therefore for each band, Chern number satisfies the following equation: $C(U)=-C(U)$ which naturally implies that $C(U=0)=0$.

\subsubsection{Type II BG/BG}

There is a different system setup: AB stacked BG on BA stacked BG. To model this different system, we label $\psi_t=\left(\begin{array}{c}c^t_{A_1}\\c^t_{B_1}\\c^t_{A_2}\\c^t_{B_2}\end{array}\right)$ for top layer and  $\psi_b=\left(\begin{array}{c}c^b_{B_1}\\c^b_{A_1}\\c^b_{B_2}\\c^b_{A_2}\end{array}\right)$ for bottom layer. 

The Hamiltonian of valley $+$ for this  system is:
\begin{align}
  H&=\sum_{a,\mathbf k} \psi_a^\dagger(\mathbf k)H^0_a(\mathbf k)\psi_a(\mathbf k)\notag\\
  &+\sum_{\mathbf k,\mathbf {Q_j}}\left( \psi^\dagger_t(\mathbf k) T_j\psi_b(\mathbf{k+Q_j})+h.c.\right)
\end{align}

$H^0_t$ should be the same as AB stacked BG on AB stacked BG case, while $H^0_b$ should change to

\begin{align}
&H^0_b(\mathbf k)\notag\\
&=\left(
\begin{array}{cccc}
U^b_1 &\upsilon(k_x+i k_y) &0 & t_3(k_x-i k_y)\\
\upsilon(k_x-i k_y)&U^b_1 &\gamma_1 &0\\
0&\gamma_1 &U^b_2&\upsilon(k_x+i k_y)\\
t_3(k_x+ik_y)&0&\upsilon(k_x-ik_y)&U^b_2
\end{array}
\right)
\end{align}

The moir\'e hopping matrix $T_j$ also should change to
\begin{equation}
  T_j=\left(
\begin{array}{cccc}
0 &0 &0 & 0\\
0&0 &0 & 0\\
t_M e^{-ij\varphi}&t_M  &0&0\\
t_M &t_Me^{ij\varphi}&0&0
\end{array}
\right)
\end{equation}
where $\varphi=\frac{2\pi}{3}$ and $j=0,1,2$. 

One can check that this Hamiltonian is invariant under  $C_3$ rotation with $\mathbf k\rightarrow C_3 \mathbf k$, and
\begin{align}
c^t_{A_1}(\mathbf k) &\rightarrow c^t_{A_1}(C_3 \mathbf{k}) e^{i \varphi} \notag\\
c^t_{B_1}(\mathbf k) &\rightarrow c^t_{B_1}(C_3 \mathbf{k}) e^{-i \varphi} \notag\\
c^t_{A_2}(\mathbf k) &\rightarrow c^t_{A_2}(C_3 \mathbf{k}) e^{-i \varphi} \notag\\
c^t_{B_2}(\mathbf k) &\rightarrow c^t_{B_2}(C_3 \mathbf{k})  \notag\\
c^b_{A_1}(\mathbf k) &\rightarrow c^b_{A_1}(C_3 \mathbf{k}) e^{- i \varphi} \notag\\
c^b_{B_1}(\mathbf k) &\rightarrow c^b_{B_1}(C_3 \mathbf{k})  \notag\\
c^b_{A_2}(\mathbf k) &\rightarrow c^b_{A_2}(C_3 \mathbf{k}) e^{i \varphi} \notag\\
c^b_{B_2}(\mathbf k) &\rightarrow c^b_{B_2}(C_3 \mathbf{k}) e^{-i \varphi} \notag\\
\end{align}

For AB on BA BG/BG, there is a mirror symmetry $M_y$ which flips the valley index.

The transformation is:
\begin{equation}
\begin{array}{c}
c_{A_1}^t(k)\rightarrow \tilde{c}_{A_2}^b(M_y k)\\
c_{B_1}^t(k)\rightarrow \tilde{c}_{B_2}^b(M_y k)\\
c_{A_2}^t(k)\rightarrow \tilde{c}_{A_1}^b(M_y k)\\
c_{B_2}^t(k)\rightarrow \tilde{c}_{B_1}^b(M_y k),
\end{array}
\end{equation}
where $\tilde{c}$ is for another valley in the large Brillouin zone.  $M_y (k^t_x, k^t_y)= (-\tilde k^b_x, \tilde k^b_y)$, where $\tilde k^b$ is defined as relative to the band touching point of valley $-$ for bottom graphene at large BZ.

For terms in $T_j$, we have,
\begin{equation}
\begin{array}{c}
c_{A_2}^{t\dag}(k)c_{A_1}^b(k+Q_j)\rightarrow \tilde{c}_{A_1}^{b\dag}(M_y k)\tilde{c}_{A_2}^t(M_y(k+Q_j))\\
c_{B_2}^{t\dag}(k)c_{B_1}^b(k+Q_j)\rightarrow \tilde{c}_{B_1}^{b\dag}(M_y k)\tilde{c}_{B_2}^t(M_y(k+Q_j))\\
c_{A_2}^{t\dag}(k)c_{B_1}^b(k+Q_j)e^{-ij\varphi}\rightarrow \tilde{c}_{A_1}^{b\dag}(M_y k)\tilde{c}_{B_2}^t(M_y(k+Q_j))e^{-ij\varphi}\\
c_{B_2}^{t\dag}(k)c_{A_1}^b(k+Q_j)e^{ij\varphi}\rightarrow \tilde{c}_{B_1}^{b\dag}(M_y k)\tilde{c}_{A_2}^t(M_y(k+Q_j))e^{ij\varphi},
\end{array}
\end{equation}
where $M_y Q_{0,1,2}=Q_{0,2,1}$. Note that the matrix element for another valley is the time reversal partner of the original valley. For example $\tilde{c}_{A_1}^{b\dag}(k)\tilde{c}_{B_2}^t(k+Q_j))e^{ij\varphi}$. The Hamiltonian $H(k)\rightarrow\tilde{H}(M_y k)$. $\tilde{H}$ is the Hamiltonian of another valley.

Combination of $M_y$ and time reversal $T$ gives an anti-unitary symmetry which preserves valley index. Because we have $SU(2)$ spin rotation symmetry, time reversal can be implemented as simply  complex conjugation without flipping the spin. Therefore we have an anti-unitary $M_y T$ symmetry which maps $(k^t_x,k^t_y)$ to $(k^b_x,-k^b_y)$. 

$M_y T$ does not flip the sign of Berry curvature. It also reverses the direction of perpendicular electric field. Therefore we have requirement of Chern number of each band: $C(U)=C(-U)$. As a result, there is no constraint for Chern number at $U=0$ and an arbitrary Chern number is possible at $U=0$. 

\subsubsection{Magic angle for BG/BG }
Within the continuum model, it is known that the twisted bilayer graphene system has magic angles where velocity of Dirac cone is zero\cite{bistritzer2011moire}. In this section we follow Ref.~\onlinecite{bistritzer2011moire} to derive an analytical expression for the renormalized mass at band touching point with twist angle $\theta$ for BG/BG system. We assume that the stacking pattern is AB on AB.

For an isolated A-B stacked bilayer graphene, the Hamiltonian of one valley is:
\begin{equation}
h_0(k,\theta)=
\left(
  \begin{array}{cccc}
  0 & t_3k_{\theta}^* & 0 & -vk_{\theta} \\
 t_3k_{\theta} & 0 & -vk_{\theta}^* & 0\\
  0 & -vk_{\theta} & 0 & \gamma_1\\
  -vk_{\theta}^* & 0 & \gamma_1 & 0
  \end{array}
\right)
\end{equation}
with basis $(c_{A1}\ c_{B2}\ c_{A2}\ c_{B1})^T$ and $k_{\theta}=(k_x+ik_y)e^{-i\theta/2}$. Subscript $1,2$ are for layer 1 and 2, $A,B$ are for different sub-lattices. For layer 3 and 4, replace 1 by 3, 2 by 4. Here $\gamma_1$ is the hopping term between $B_1$ and $A_2$.

We consider bilayer 1,2 on top of bilayer 3,4 and hopping only between layer 2 and 3. The hopping matrix is,
\begin{equation}
T_{32}(k+Q_j,k)=t_M (c_{A3}^\dag(k+Q_j)\ c_{B3}^\dag(k+Q_j))
\left(
\begin{array}{cc}
1 & e^{ij\phi}\\
e^{-ij\phi} &1
\end{array}
\right)
\left(
\begin{array}{c}
c_{A2}(k) \\
c_{B2}(k)
\end{array}
\right)
\end{equation}
$Q_j$ is moir\'e reciprocal lattice vector, $Q_0=
K_\theta(0,-1)$, $Q_1=K_\theta(\sqrt{3}/2,1/2)$, and $Q_2=K_\theta(-\sqrt{3}/2,1/2)$.

Integrating the higher energy orbitals $B_1,A_2,B_3,A_4$, the effective Hamiltonian is,
\begin{equation}
h_{eff}(k+Q_j,k)=
\left(
  \begin{array}{cccc}
  0 & t_3k_{\theta}^*-v^2k_{\theta}^2/\gamma_1 & t_Mvk_{\theta}/\gamma_1 & t_Mv^2k_{\theta}(k+Q_j)_{-\theta}e^{ij\phi}/\gamma_1^2 \\
 c.c & 0 & t_Me^{-ij\phi} & t_Mv(k+Q_j)_{-\theta}/\gamma_1\\
 c.c & c.c & 0 &t_3(k+Q_j)_{-\theta}^*-v^2(k+Q_j)_{-\theta}^2/\gamma_1\\
c.c &c.c &c.c& 0
  \end{array}
\right)
\end{equation}
where $c.c$ stands for complex conjugate. The basis is $(c_{A1}\ c_{B2}\ c_{A3}\ c_{B4})^T$.

We can write down an 8-band model to gain some insight. We focus on the band touching point  for valley $+$ of the top graphene layer, which is at $K$ point of the MBZ. In the limit $k^t\rightarrow 0$, each momentum $k^t$ connects to three momentums of the bottom graphene  layer: $k^b=k^t+Q_j$. Keeping these four momentum points, we can write an eight band model. 

At small $|k|$,  $H(k)=H^{(0)}+H^{(1)}(k)$, where,
\begin{equation}
H^{(0)}=
\left(
  \begin{array}{cccc}
  h_0 & T_1 & T_2 &T_3 \\
 T_1^\dag & h_1 & 0 & 0\\
 T_2^\dag & 0 & h_2 & 0\\
T_3^\dag & 0 & 0& h_3
  \end{array}
\right),
\end{equation}
$h_j$, $T_j$ are $2\times 2$ matrix, for $j=1,2,3$,
\begin{equation}
h_j=\left(
  \begin{array}{cc}
  0 & -\frac{v^2}{\gamma_1} (Q_j)_{-\theta}^2\\
 -\frac{v^2}{\gamma_1}((Q_j)_{-\theta}^2)^* &0
  \end{array}
\right)
\end{equation}
\begin{equation}
T_j=\left(
  \begin{array}{cc}
  0 & 0\\
t_M e^{-ij\phi} &\frac{t_Mv}{\gamma_1}(Q_j)_{-\theta}
  \end{array}
\right)
\end{equation}
The zero-eigenvalue states can be written as $\psi=(\psi_0,\psi_1,\psi_2,\psi_3)$, where each entry is a 2-component spinor. We can obtain the wavefunction of the zero energy state as:
\begin{equation}
\psi_j=-h_j^{-1}T_j^\dag\psi_0=\frac{t_M}{vK_\theta^4}\left(
  \begin{array}{c}
   K_\theta^2 (Q_j)_\theta\\
\frac{\gamma_1}{v}e^{ij\phi} (Q_j)_\theta^{2*}
  \end{array}
\right)\psi_0^{(2)},
\end{equation}
and the square-norm of $\psi$ is,
\begin{equation}
|\psi|^2=\psi^\dag_0\left(
  \begin{array}{cc}
  1 & 0\\
0 &1+3\frac{\gamma_1^2t_M^2}{v^4K_\theta^4}(1+\frac{v^2K_{\theta}^2}{\gamma_1^2})
  \end{array}
\right)\psi_0
\end{equation}

Note that $\psi_0$ forms a two dimensional Hilbert space. Therefore $\psi$ also forms a two dimensional space.

The block between $k$ and $k+Q_j$ in the perturbation term $H^{(1)}(k)$ is,
\begin{equation}
H^{(1)}(k+Q_j;k)=\left(
  \begin{array}{cccc}
  0 & -v^2k_{\theta}^2/\gamma_1 & t_M v k_\theta/\gamma_1 & t_M (v k_\theta/\gamma_1)(v(k+Q_j)_{-\theta}/\gamma_1)  e^{ij\phi} \\
 c.c & 0 & 0 & t_Mv k_{-\theta}/\gamma_1\\
 c.c & c.c & 0 & -v^2k_{-\theta}(k+2Q_j)_{-\theta}/\gamma_1\\
c.c &c.c &c.c& 0
  \end{array}
\right),
\end{equation}
where $j=1,...,3$, $H^{(1)}(k)$ is an $8\times 8$ matrix.

We can obtain the elements of $H^{(1)}(k)$ in the $\psi$ basis, where the $\theta$ dependence of wave vectors are ignored,
\begin{eqnarray}
H^{(1)}(k)_{11}&=&0\\
H^{(1)}(k)_{12}&=&-\frac{v^2}{\gamma_1}(k_x+ik_y)^2(1-3\frac{t_M^2}{v^2K_\theta^2})/\alpha\\
H^{(1)}(k)_{22}&=&0
\end{eqnarray}
where the normalization factor $\alpha=\sqrt{1+3\frac{\gamma_1^2t_M^2}{v^4K_\theta^4}(1+\frac{v^2K_{\theta}^2}{\gamma_1^2})}$.

One can see that the mass of the original quadratic band is enhanced by a factor $m^*=\alpha m_0/(1-3\frac{t_M^2}{v^2K_\theta^2})$.

Interestingly $m^* \rightarrow \infty$ at magic angle decided by $1-3\frac{t_M^2}{v^2K_\theta^2}=0$. This equation gives $\theta_M=1.08^\circ$ which is equal to the magic angle value of twisted monolayer graphene on monolayer graphene system\cite{bistritzer2011moire}.

\subsection{Analysis of Chern number}

In the following we give some insights about the phase diagram of Chern number for twisted graphene on graphene systems.

\subsubsection{Type I: Double graphenes with opposite chirality}

 For either BG or TG, we can only keep two bands. We label the number of layers for top graphene as $n_t$ and the number of layers for bottom graphene as $n_b$. The Hamiltonian of each graphene is:
\begin{equation}
  H_a=\psi^\dagger(k)
  \left(
  \begin{array}{cc}
  m_a & \frac{\upsilon^{n_a}}{\gamma_1^{n_a-1}}(k_x-ik_y)^{n_a} \\
  \frac{\upsilon^{n_a}}{\gamma_1^{n_a-1}}(k_x+ik_y)^{n_a} & -m_a
  \end{array}
  \right)
  \psi(k)
\end{equation}
where $\psi_a=\left(\begin{array}{c}c^a_1\\c^a_2\end{array}\right)$. $c^a_1$ and $c^a_2$ are the first layer and the second layer for graphene $a=t,b$. Layer $1$ is always fixed to be on top of layer $2$. Therefore mass $m_a$ is proportional to energy difference between top layer and bottom layer, up to an offset.  We put $\psi_t$ at $K_M$ of MBZ and $\psi_b$ at $K'_M$ of MBZ. Momentum $\mathbf k$ in the above hamiltonian is defined relative to $K_M$ or $K'_M$.

Moir\'e hopping Hamiltonian is:
\begin{equation}
H_M=\sum_{\mathbf k,\mathbf {Q_j}}\left(t_M e^{i j \varphi} c^{t\dagger}_2(\mathbf k) c^b_1(\mathbf{k+Q_j}) +h.c.\right)
\end{equation}
where $\varphi =\frac{2\pi}{3}$. 

We have projected moir\'e hopping matrix in Eq.~\ref{eq:tM_AB_AB} to low energy band $\psi_a$ here.  

There are also term from higher order perturbations when we integrate higher energy degree of freedoms $c^t_{B_1}$, $c^t_{A_2}$, $c^b_{B_1}$ and $c^b_{A_2}$. The most important term is the following mass term $A c^{t\dagger}_2(k)c^t_2(k)+A c^{b\dagger}_1(k)c^b_1(k)$. Therefore at $U=0$, we have mass term $m_t$ and $m_b$ with opposite sign.  For BG/BG and TG/TG system, $M_x$ symmetry requires $m_t=-m_b$. These mass terms gap out quadratic band touching  at $K$ and $K'$ with opposite sign. 

At $U=0$, contributions to Chern number from $K_M$ and $K'_M$ are $C_{K}=\frac{n_t}{2}$, $C_{K'}=-\frac{n_b}{2}$. For BG/BG and TG/TG system, at $U=0$, Chern number is $C(U=0)=0$ as required by $M_x$ symmetry. For  TG/BG system, we find that $C(U=0)=-1$, resulting from an additional contribution at $\Gamma$ point.

Next we can easily understand Chern numbers at non-zero $|U|$. When $U>\Delta_0>0$, $m_b$ changes to positive through a gap closing and reopening process. It is easy to figure out that Chern number of the conduction band need to change by $-n_b$.  Therefore $C(U>\Delta_0)=-n_b$ for TG/TG and BG/BG system. For TG/BG system, $C(U>\Delta_0)=-1-n_b=-3$.

Similarly, when $U<-\Delta_0$, $m_t$ changes to negative and total Chern number of the conduction band needs to change by $n_t$. Therefore $C(U<-\Delta_0)=n_t$ for TG/TG and BG/BG system. For TG/BG, $C(U<-\Delta_0)=n_t-1=2$.

When we further increase $|U|$, the hybridization gap between the conduction band and the band above closes at $U_c$ and reopens after $|U|>U_c$. Through this process Chern number $|C|$ drops by $1$. $U_c(\theta)$ vanishes at magic angle $\theta_M$.

\subsubsection{Type II: Double graphenes with opposite chirality}
When the top and bottom graphenes have the opposite chirality, the Hamiltonians for them are:
\begin{equation}
  H_t=\psi^\dagger(k)
  \left(
  \begin{array}{cc}
  m_t & \frac{\upsilon^{n_t}}{\gamma_1^{n_t-1}}(k_x-ik_y)^{n_t} \\
  \frac{\upsilon^{n_t}}{\gamma_1^{n_t-1}}(k_x+ik_y)^{n_t} & -m_t
  \end{array}
  \right)
  \psi(k)
\end{equation}
and
\begin{equation}
  H_b=\psi^\dagger(k)
  \left(
  \begin{array}{cc}
  m_b & \frac{\upsilon^{n_b}}{\gamma_1^{n_b-1}}(k_x+ik_y)^{n_b} \\
  \frac{\upsilon^{n_b}}{\gamma_1^{n_b-1}}(k_x-ik_y)^{n_b} & -m_b
  \end{array}
  \right)
  \psi(k)
\end{equation}
We still put $\psi_t$ on $K$ and $\psi_b$ on $K'$ of MBZ.

The moir\'e hopping Hamiltonian is:

\begin{equation}
H_M=\sum_{\mathbf k,\mathbf {Q_j}}\left(t_M c^{t\dagger}_2(\mathbf k) c^b_1(\mathbf{k+Q_j})+h.c.\right)
\end{equation}
where $\varphi =\frac{2\pi}{3}$. 

We also simulate the above Hamiltonian numerically. 

At $U=0$, we still have $m_t$ and $m_b$ with opposite sign. However, quadratic band touching at $K$ and $K'$ have the opposite chirality. Therefore contributions to Chern number are $C_K=\frac{n_t}{2}$, $C_{K'}=\frac{n_b}{2}$. This is also consistent with symmetry requirements for BG/BG and TG/TG system, for which $M_yT$ symmetry requires that $K$ and $K'$ contribute the same berry curvature. The contribution from region close to $\Gamma$ point is arbitrary. Therefore there is no constraint for Chern number at $U=0$. We have $C(U=0)=C_0$ with $C_0$ a general integer which depends on the system and twist angle.

When $U>\Delta_0$, $m_b$ becomes positive and Chern number should change by $n_b$. Therefore conduction band $C(U>U_c)=C_0+n_b$.

When $U<-\Delta_0$, $m_t$ changes to negative and Chern number changes by $n_t$. Therefore  conduction band $C(U<-U_c)=C_0+n_t$.

The change of Chern number for the valence band across $|\Delta_0|$ is opposite to the conduction band, although $C_0$ of conduction and valence bands have no obvious correlation.

When further increasing $|U|$, the hybridization gap between conduction band and the band above closes at $U_c$ and reopens, resulting a further drop of Chern number by $1$, similar to type I system. $U_c(\theta)$ also vanishes at the magic angle.

\section{Skyrmion Energy of IQAH Insulator \label{appendix:skyrmion}}
We try to estimate the energy cost of skyrmion excitation of the spin and valley polarized integer QAH state at total filling $\nu_T=1$ or $\nu_T=3$. We focus on the $C=2$ case, though the following calculation can be easily generalized to other Chern number.

We assume the band filled is $(-,\downarrow)$, in the following we ignore valley index because we expect valley flip excitation is not important. The ground state wavefunction is:
\begin{equation}
	\ket{\Omega}=\prod_{\mathbf k}c^\dagger_\downarrow (\mathbf k)\ket{0}
\end{equation}

To estimate the energy cost of skyrmion, we should estimate spin phase stiffness $\rho_s$ first. This can be done by calculating the dispersion of spin wave.  Spin wave with momentum $\mathbf q$ is created by

\begin{equation}
	S^\dagger(\mathbf q)=\sum_{\mathbf k}c^\dagger_\uparrow(\mathbf k +\mathbf q)c_{\downarrow}(\mathbf k) \lambda(\mathbf k, \mathbf{k+q}) 
\end{equation}
where we still need to include the form factor because we're projecting microscopic spin operator to the Chern band.

Again we only consider interaction energy:
\begin{equation}
	H_V=\sum_{\mathbf q} V(|\mathbf q|)\tilde \rho(\mathbf q) \tilde \rho(-\mathbf q)
\end{equation}
where we only consider density operator of valley $-$.

To make analytical calculation possible, we assume Berry curvature is uniform in BZ and then we can write the form factor in the following form:
\begin{equation}
	\lambda(\mathbf k, \mathbf {k+q})=F(|\mathbf q|)e^{\frac{B}{2}i\mathbf k \wedge \mathbf q}
\end{equation}
where $F(|q|)$ is a real function which decays with $|\mathbf q|$.  $B=C\frac{2\pi}{A_{BZ}}$ is the uniform Berry curvature.

Then it is easy to get the following commutation relation:
\begin{equation}
	[\tilde \rho(\mathbf{q_1}),\tilde \rho(\mathbf{q_2})]=\frac{F(|\mathbf q_1|)F(|\mathbf q_2|)}{F(|\mathbf{q_1+q_2}|)}2i\sin(B \frac{\mathbf{q_1} \wedge \mathbf{q_2}}{2}) \tilde \rho(\mathbf{q_1}+\mathbf{q_2})
\end{equation}

\begin{equation}
	[\tilde \rho(\mathbf{q_1}),S^\dagger(\mathbf{q_2})]=\frac{F(|\mathbf q_1|)F(|\mathbf q_2|)}{F(|\mathbf{q_1+q_2}|)}2i\sin(B \frac{\mathbf{q_1} \wedge \mathbf{q_2}}{2}) S^\dagger(\mathbf{q_1}+\mathbf{q_2})
\end{equation}
which resemble the GMP algebra in Projected Landau Level except an additional factor $B$.

Then it is easy to calculate spin wave dispersion following the standard approach.

\begin{equation}
	[H_V,S^\dagger(\mathbf q)]\ket{\Omega}=4 \sum_{\mathbf p} \sin^2(B \frac{\mathbf p \wedge \mathbf q}{2}) F^2(|\mathbf p|)V(|\mathbf p|)  S^\dagger(\mathbf q)\ket{\Omega}
\end{equation}
During this derivation we use the fact that $\rho(\mathbf q)\ket{\Omega}=0$ for any $\mathbf q \neq 0$.

Therefore the dispersion of spin-wave excitation is:
\begin{equation}
	\xi(\mathbf q)=4 \sum_{\mathbf p} \sin^2(B \frac{\mathbf p \wedge \mathbf q}{2}) V(|\mathbf p|) F^2(|\mathbf p|)
\end{equation}

At $|\mathbf q|\rightarrow 0$ limit, we get dispersion of spin wave:
\begin{equation}
	\xi(\mathbf q)=A |\mathbf q|^2
\end{equation}
with
\begin{equation}
	A=\frac{B^2}{2} \sum_{\mathbf p} |\mathbf p|^2 V(|\mathbf p|) F^2(|\mathbf p|)
	\label{eq:phase_stiffness}
\end{equation}
where we used the fact that $\langle \sin^2 \theta \rangle=\frac{1}{2}$ when integrating over $\theta \in [0,2\pi)$.

From the spin wave dispersion we can easily extract the spin phase stiffness and then estimate the energy of skyrmion. Following Ref.~\cite{girvin1999quantum}, the phase stiffness is
\begin{equation}
	\rho_s=\frac{n  B^2}{4} \sum_{\mathbf p} |\mathbf p|^2 V(|\mathbf p|) F^2(|\mathbf p|)
\end{equation}
 We can define length in the unit of $a_M$ and momentum in the unit of $\frac{1}{a_M}$. Therefore density of electrons is $n=1$.  There is a $\frac{1}{2}$ factor coming from spin $S=\frac{1}{2}$. In the following we will keep $n$ to make the dimension of  the equation obvious.

Energy of skyrmion is
\begin{equation}
	E_{skyrmion}=4\pi \rho_s= n \pi B^2 \sum_{\mathbf p} |\mathbf p|^2 V(|\mathbf p|) F^2(|\mathbf p|)
\end{equation}

and energy of widely separated skyrmion-antiskyrmion pair is

\begin{equation}
	E_{skyrmion pair}=2\pi n B^2 \sum_{\mathbf p} |\mathbf p|^2 V(|\mathbf p|) F^2(|\mathbf p|)
	\label{eq:skyrmion-antiskyrmion}
\end{equation}

When $|\mathbf q|$ is large, essentially $S^\dagger(\mathbf q)=\sum_{\mathbf k}c^\dagger_{\uparrow}(\mathbf{k+q})c_{\downarrow}(\mathbf k)\lambda(\mathbf{k},\mathbf{k+q})$ creates an electron-hole pair with total momentum $\mathbf{q}$. Because of GMP algebra implicitly hidden in $\lambda(\mathbf k, \mathbf{k+q})$, the distance $r$ of this electron-hole pair is proportional to $|\mathbf q|$.  $\xi(|\mathbf q|\rightarrow \infty)$ actually corresponds to energy cost of infinitely separated charge $e$ particle-hole pair.
\begin{equation}
	E_{1e;PH}=2\sum_{\mathbf p}  V(|\mathbf p|) F^2(|\mathbf p|)
\end{equation}
where we approximate $\sin^2$ term with its averaged value $\frac{1}{2}$.

Correspondingly energy cost of charge $2e$ particle-hole pair should be:
\begin{equation}
	E_{2e;PH}=4\sum_{\mathbf p}  V(|\mathbf p|) F^2(|\mathbf p|)
	\label{eq:2ePH}
\end{equation}

The energy cost is apparently sensitive to the form of $F(|\mathbf p|)$, which can only be calculated numerically.

To gain some intuition, we assume the following  form of $F(|\mathbf p|)$: $F^2(|\mathbf p|)=e^{-\frac{\alpha}{2}|\mathbf p|^2}$. $V(|\mathbf p|)=\frac{1}{|\mathbf p|}$ is Coulomb interaction. Then it is easy to get:
\begin{equation}
	\frac{E_{skyrmion pair}}{E_{2e;PH}}=\frac{\pi n B^2}{2 \alpha}
\end{equation}
where we add electron density $n$ to make the above quantity dimensionless.

For classic $\nu_T=1$ Lowest Landau Level problem, we have $\alpha=1$, $B=1$, $n=\frac{1}{2\pi}$ for which the fundamental unit of length is $l_B$.  $\frac{E_{skyrmion pair}}{E_{2e;PH}}=\frac{1}{4}$ is in agreement with the well-known fact that $\frac{E_{skyrmion pair}}{E_{1e;PH}}=\frac{1}{2}$ \cite{sondhi1993skyrmions}.

For the systems we studied in this paper, we use $a_M$ as the unit of length and then we have Hexagonal BZ with $A_{BZ}=\frac{8\pi^2}{\sqrt{3}}$ and density $n=1$. Then the uniform Berry curvature is $B=\frac{2\pi C}{A_{BZ}}=\frac{\sqrt{3}}{4\pi} C$. Therefore for $C=2$:
\begin{equation}
	\frac{E_{skyrmion pair}}{E_{2e;PH}}=\frac{3 }{8 \pi \alpha}
\end{equation}
Therefore when $\alpha>\frac{3}{8\pi}$, Skyrmion is the cheapest charged excitation.

For the moir\'e super-lattice system, we can numerically calculate $F^2(|\mathbf q|)=|\lambda(0,\mathbf q)|^2$.  We can also numerically do the integration in Eq.~\ref{eq:skyrmion-antiskyrmion} and Eq.~\ref{eq:phase_stiffness} directly.

For valence band of  BG/h-BN at $\theta=0$,  we get $\frac{E_{skyrmion pair}}{E_{2e;PH}}=0.58$ by directly integrating Eq.~\ref{eq:skyrmion-antiskyrmion} and Eq.~\ref{eq:phase_stiffness}. Therefore we expect finite densities of skyrmions emerge even in the ground state upon doping away from $\nu_T=1$.

For conduction band of type I BG/BG,  at $\theta=1.17^\circ$ and $U=8$ meV, we get $\frac{E_{skyrmion pair}}{E_{2e;PH}}=1.35$. At $\theta=0.9^\circ$ and $U=8$ meV, $\frac{E_{skyrmion pair}}{E_{2e;PH}}=0.87$. Therefore whether skyrmions emerge upon doping is quite sensitive to twist angle for this system. Smaller twist angle may be favorable to search for skyrmion superconductor.

\section{Fractional Quantum Valley Hall Effect \label{appendix:FQVHE}}
We focus on filling $\nu_T=2+\nu_{eff}$. As we argued in the main text, $\nu_T=2$ is a spin polarized QVH state. Then if we dope additional $\nu_{eff}$ electrons, they must carry the other spin.  Therefore we expect to have spin polarized topological ordered state with effective filling $\nu_{eff}=\nu_T-2$.

One option is that valley is also polarized and we get a FQAH state with effective filling $\nu_{eff}$. However, it may also be possible to have a state for which two valleys have equal density. This state is likely to be Fractional Quantum Valley Hall(FQVH) state with different valleys having opposite hall conductivity.   Based on  our discussion of FQAH state, there could be spin polarized FQVH state at filling $\nu_{eff}=\frac{2m}{2km|C|+1}$. These states have zero charge hall conductivity and valley hall conductivity $\sigma^\upsilon_{xy}=\nu_{eff}C$.

Taking $|C|=1$ as an example, we can have a FQVH state at effective filling $\nu_{eff}=\frac{2}{3}$, which is a $\frac{1}{3}$ Laughlin state  for one valley plus  $-\frac{1}{3}$ Laughlin state for another valley. For $|C|>1$ bands, we can have similar states.

One interesting direction to explore in the future is to construct new topological ordered state beyond two decoupled FQHE states with opposite Hall conductivities.

\end{document}